\documentclass[prb,aps,twocolumn,showpacs]{revtex4-1}

\usepackage{graphicx}
\usepackage{bm,color}
\usepackage{physics}
\usepackage{amsmath, amssymb}
\usepackage{bm,color}
\usepackage{multirow}

\newcommand{\be}{\begin{eqnarray}}
\newcommand{\ee}{\end{eqnarray}}

\usepackage{color}
\usepackage{ulem}

\def\l{\langle}

\def\s{\sigma}
\def\d{\delta}
\def\HIC{H_{\rm IC}}
\def\HCS{H_{\rm CS}}
\def\l{\lambda}
\def\ua{\uparrow}
\def\da{\downarrow}

\def\Et{EE(t)}

\begin{document}

\title{Quantum information spreading in  random spin chains with topological order
}
\date{\today}
\author{Takahiro Orito$^{1}$}
\author{Yoshihito Kuno$^{2}$}
\author{Ikuo Ichinose$^{3}$}

\affiliation{$^1$Graduate School of Advanced Science and Engineering, Hiroshima University, 739-8530, Japan}
\affiliation{$^2$Graduate School of Engineering Science, Akita University, Akita 010-8502, Japan}
\affiliation{$^3$Department of Applied Physics, Nagoya Institute of Technology, Nagoya, 466-8555, Japan}

\begin{abstract}
Quantum information spreading and scrambling in many-body systems attract interests 
these days.
Tripartite mutual information (TMI) based on operator-based entanglement entropy (EE) is an efficient tool for measuring them.
In this paper, we study random spin chains that exhibit phase transitions accompanying 
nontrivial change in topological properties. 
In their phase diagrams, there are two types of many-body localized (MBL) states and 
one thermalized regime intervening these two MBL states.
Quench dynamics of the EE and TMI display interesting behaviors providing
essential perspective concerning encoding of quantum information.
In particular, one of the models is self-dual, but information spreading measured by 
the TMI does not respect this self-duality.
We investigate this phenomenon from the viewpoint of spatial structure of the 
stabilizers. 
In general, we find that knowledge of phase diagram corresponding to qubit system 
is useful for understanding nature of quantum information spreading in that system.
Connection between the present work and random circuit of projective measurements 
and also topological Majorana quantum memory is remarked.
\end{abstract}
\maketitle


\section{Introduction}

Study of many body localization (MBL) is one of central issues in the condensed matter physics \cite{Nandkishore2015,Abanin2017,Alet,Abanin2019}. 
In particular, the entanglement property in the bulk for excited states has been
investigated  
and it is verified that its system-size scaling law behaves differently from that of 
the thermal phase \cite{Alet,Khemani2017}. 
Also, MBL exhibits a non-trivial dynamical aspect for the quenching time evolution 
of entanglement entropy (EE). 
In typical MBL phases, for a specific type of an initial state, the EE exhibits a logarithmic growth, due to the presence of interactions, where particles (or spin) are not transported
but the EE gradually spreads into an entire system \cite{Znidaric,Bardarson}. 
So far, various types of the MBL states have been proposed through lots of theoretical and 
numerical works, such as spin-glass MBL (SG-MBL) \cite{Huse2013,Kajall2014}, MBL induced by quasi-periodic potential \cite{Iyer2013},
topological MBL \cite{Huse2013,Bahri,Decker2020,Sahay2021}, MBL emerging in lattice gauge theories \cite{Smith2017,Smith2018,Park2019} and 
various disorder-free MBLs \cite{Schulz2019,Nieuwenburg2019,OKI2020,KOI2020,Danieli_1,Roy,Zurita,Danieli_2,Danieli_3,OKI2021,OKI2021_2}, etc. 
However, it is expected that there are still many different types of MBL categories,
which have not been explored yet.
Moreover, the unique spectral and entanglement structures associated with each 
type of MBL are diverse and study of these detailed structures gives us an insight to
understand how quantum information is encoded and stored in MBL regimes.
Study of these issues is useful for understanding deeply encoding and storing mechanism of
quantum information in localization systems. 

In this work, we study two disordered spin models exhibiting characteristic
multiple topological MBL phases. 
The disordered spin models exhibit rich phase diagrams because of the emergence of different sets of effective stabilizers for each phase, local integrals of motion (LIOMs) \cite{Nandkishore2015} in the context of MBL, and these different sets of stabilizers are 
non-commutative with each other. 
The stabilizers in the disordered models respect symmetries of the models and become basic building blocks of the topological order \cite{Briegel_2001,Son2011,Smacchia,Bahri,Decker2020,Wahl}. 
Furthermore, the spatial structure of the stabilizers influences the bulk entanglement
structure in the MBL phases and also degeneracy of energy spectrum in the whole band \cite{Bahri,Decker2020}
(related to the presence of gapless edge modes).

One of our target spin models, random transverse field Ising model \textit{at infinite
temperature}, was extensively studied
recently~\cite{Laflorencie2022,Wahl2022,Sahay2021,Moudgalya2020}. 
There, the global phase diagram, which includes two types of the MBL phases 
(paramagnetic MBL and SG-MBL phases), was obtained by numerical
investigation.
In this work, we shall study detailed properties of these MBL phases such as quenching
dynamics of the EE from the viewpoint of duality. 
Then we employ some quantum information theoretic quantity, tripartite mutual
information (TMI) proposed in Ref.~\onlinecite{Hosur}, to investigate the topological MBL
from the viewpoint of information spreading. 
As a measure of the scrambling, the out-of-time-ordered correlator (OTOC) was
proposed~\cite{Shenker2014,Maldacena2016}, and it was applied to some kind of quantum spin 
models~\cite{Swingle2017,He2017,Sahu2019}.
Compared with the OTOC, the TMI is state and operator independent, and it is becoming
a benchmark of the quantum information spreading nowadays.
In this work, we shall numerically demonstrate that the system-size dependence of 
the TMI is valid to identify phase boundaries of the system. 
Furthermore, we shall study another disordered spin model having two different types of
topological MBL in its phase diagram.
We clarify the model's global phase structure by varying the strength of two types of
disorders, and observe the bulk information spreading in the whole parameter region. 
In particular, we show that the bulk structure of the information spreading in the
topological MBLs is captured by using the TMI and also is determined by the
spatial structure of the stabilizers in each phase of MBL.

Among the findings in this work obtained by the numerical calculation, an interesting
observation concerns the infinite random criticality (IRC) and Griffiths phase \cite{Fisher1995,Young1996,Fisher1999} in the random transverse field Ising spin chain.
This model has been studied for a long time as one of the most important models for understanding random systems. 
We shall shed light on its physical properties from the viewpoint of quantum information
scrambling in this work. 
As another interesting observation, by the calculation of the TMI, we acquire 
an important insight into how quantum information in the bulk is encoded in 
quantum spin chains and how disorder influences quantum information spreading.
Calculation of the TMI in the two-site partitioning of chain reveals that quantum information is encoded in stabilizer-qubits in the MBL regimes.

The rest of this paper is organized as follows.
In Sec.~II, we shall introduce our target two disordered spin models and explain the 
basic properties of them. 
We also introduce the TMI and explain its practical calculation methods briefly. 
In Sec.~III, we show the results of the numerical study by means of the exact diagonalization.
Detailed discussions on the numerical results are given there to obtain
observations explained in the above.
Section IV is devoted to discussion and conclusion.

\section{Models and Tripartite Mutual information}
In this section, we introduce two types of spin chains, and briefly study their phase diagrams.
Then, we explain the TMI and methods of the practical numerical calculation.

\subsection{Models}

The first model describes a self-dual random Ising spin chain, whose Hamiltonian is 
given as follows,
\be 
H_{\rm IC} &=& \sum_i \Big[ J_i \s^x_i \s^x_{i+1} + h_i \s^z_i\Big] + H_g,   \nonumber  \\
H_g &=& g \sum_i \Big[ \s^x_i \s^x_{i+2} +\s^z_i \s^z_{i+1} \Big],
\label{HIC}
\ee
where $\s^x_i, \s^z_i$ are Pauli matrices residing on site $i$ of the chain, $J_i$ 
and $h_i$ are random 
couplings drawn from uniform distributions $[0, W_J]$ and $[0,W_h]$, respectively, 
and $g$ is a non-negative coupling constant.
By symmetry of the free part of the Hamiltonian, $H_{\rm IC}|_{g=0}$, 
$\{J_i\}$ and $\{h_i\}$ can be transformed to positive values, 
and therefore we have chosen the above parameter region.
For the practical calculation, we set $W_J = (W_h)^{-1} =W$ and introduce a parameter 
such as 
$\d = 2\ln W = \overline{\ln W_J} -\overline{\ln W_h}$.
We are interested in the phase diagram of the system $H_{\rm IC}$ [Eq.~(\ref{HIC})]
in the $(\d-g)$ plane.

It is easily verified that the system $H_{\rm IC}$ has $\mathbb{Z}_2$ symmetry by
$\mathbb{P}\equiv \prod_i \s^z_i$, and it is also self-dual by the following duality transformation;
\be
\tau^z_i = \s^x_i \s^x_{i+1}, \; \; 
\tau^x_i=\prod_{j\leq i}\s^z_j,
\label{Dutr}
\ee  
and under Eq.~(\ref{Dutr}), $\d \to - \d$.
The above properties of $H_{\rm IC}$ play an important role in the subsequent investigation of quantum information spreading in that model. 
Also for large $W$, the model can be regarded as a projective Hamiltonian with
effective stabilizers, i.e., LIOMs in the localization literature. 
These are a set of dimers $\{\sigma^{x}_{i}\sigma^{x}_{i+1}\}$, each of which approximately
commutes with $H_{\rm IC}$, $[\sigma^{x}_{i}\sigma^{x}_{i+1},H_{\rm IC}]\approx 0$ for 
any $i$. 
The presence of the stabilizers gives an insight into the bulk property of information spreading \cite{Zeng2016}.

In this paper, we are interested in the system $H_{\rm IC}$ at infinite temperature.
Phase diagram of that system has been obtained recently \cite{Laflorencie2022}, and 
there exist three phases in the phase diagram, i.e.,
paramagnetic MBL phase (PM-MBL) for $\delta < \delta_{1c}$,  ergodic regime for
$\delta_{1c} <\d<\d_{2c}$,
and MBL phase with a spin-glass/topological order (SG-MBL) for $\d_{2c} <\d$.
Values of the criticality $\d_{1c}, \d_{2c}$ depend on the strength of the coupling $g$,
and $\d_{2c}=-\d_{1c}\equiv \d_c$ by duality.
For the pure transverse random Ising model (TRIM) at $g=0$, $\d_c=0$ showing infinite 
random criticality (IRC) as in the ground state \cite{Fisher1995}.
For the infinite-temperature system, detailed investigation of the EE and gap ratio for
small $g$ by using the system-size dependence and scaling indicates the possibility of 
an intermediate ergodic phase in a finite $\delta$-region such as 
$-\ln 2 <\delta <\ln 2$ in the limit $g \to 0$~\cite{Laflorencie2022}.
This result supports the avalanche picture of the localization-delocalization transition.
It is a very interesting problem if the avalanche picture emerges in dynamics of quantum information spreading.
We shall comment on this after studying the TMI, which is a benchmark of the scrambling.

\begin{figure}[t]
\begin{center} 
\includegraphics[width=7.0cm]{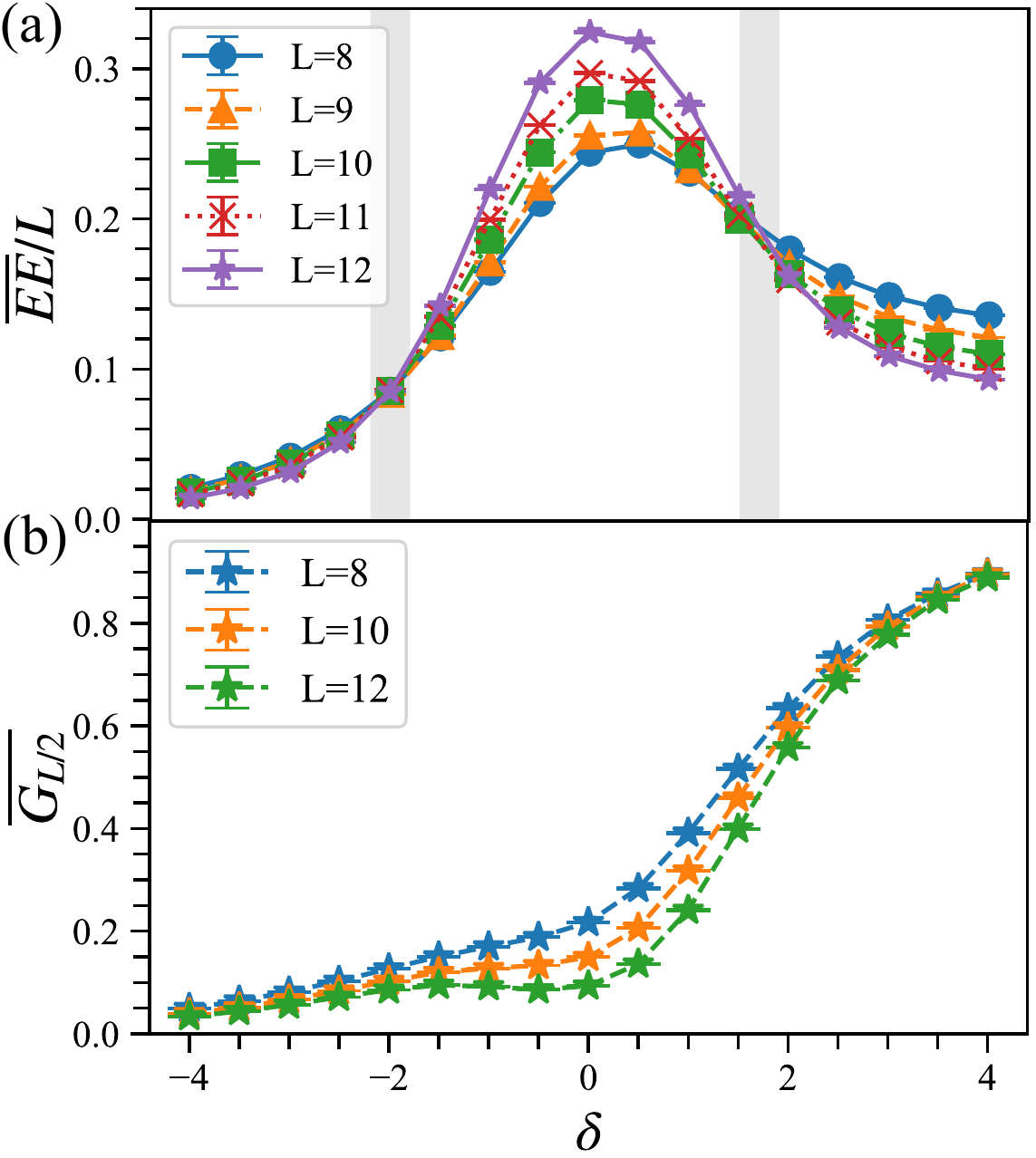}  
\end{center} 
\caption{(a) Half-chain EE for the Ising spin chain model $\HIC$ in Eq.~(\ref{HIC})
for $g = 0.2$.
Calculation of $\overline{EE}/L$ indicates the existence of two phase transitions
such as the PM-MBL $\to$ ETH $\to$ SG-MBL as $\d$ increases.
As $\d$ is getting large, $\overline{EE} \to \log 2$ corresponding to the SG-order. 
These results were obtained by averaging over $1000, 750, 500, 300,$ and $150$ disorder realization using all eigenstates for the $L = 8, 9, 10, 11, 12$ systems.
(b) Spin correlation, $G_{L/2}$ in Eq.~(\ref{Gr}), as a function of $\d$. 
Its increase indicates the SG-order for $\d \gg1$. These results were obtained by averaging over the 20000 eigenstates using 10-20 eigenstates in the middle of the spectrum at each disorder realization for the L=8, 10, and 12 systems. The error bars are standard error.
}
\label{Fig1}
\end{figure}

In Fig.~\ref{Fig1} (a), we show the numerical calculations of the half-chain EE for energy
eigenstates of $\HIC$ for $g=0.2$, which is defined as follows;
\be
&&EE^{(s)} = -\mbox{Tr}_A \biggr[\rho^{(s)}_{\rm r} \log (\rho^{(s)}_{\rm r})\biggl], \;\; 
\rho^{(s)}_{\rm r} =\mbox{Tr}_{\bar{A}}\biggr[|\psi_s\rangle\langle \psi_s| \biggl],  \nonumber \\
&& \overline{EE} = \mbox{average of $EE^{(s)}$ over states and randomness},  \nonumber
\ee
where the suffix $(s)$ denotes the combined label of sample number and state label, 
$A$ and $\bar{A}$ are the half chain and its complement, respectively.
[Hereafter, ``$\log$" denotes ``$\log_2$".]
The calculations exhibit the cirtical value $\d_c\simeq 2.0$ for $g=0.2$.
The EE has a nonvanishing value $\sim \log 2$ 
for the deep SG-MBL regime,
whereas it reduces to very small in the deep PM-MBL.
This result indicates that in the deep SG-MBL, cat states of a parity pair emerge there
such as 
$$
{1\over \sqrt{2}}(|\ua\ua\da\da\cdots\rangle\pm |\da\da\ua\ua\cdots\rangle),
$$
in the $\s^x$-basis,
and then, the reduced density matrix $\rho^{\rm sg}_{\rm R}$ is obtained as,
$$
\rho^{\rm sg}_{\rm R}= {1 \over 2} (|\ua\ua\da\da\cdots\rangle\langle \ua\ua\da\da\cdots|
+|\da\da\ua\ua\cdots\rangle\langle \da\da\ua\ua\cdots\rangle),
$$
which gives $\log 2$ for the EE.

The above observation implies the possibility that a pair of states
$(|\ua\ua\da\da\cdots\rangle,|\da\da\ua\ua\cdots\rangle)$ form a bulk qubit and quantum information is encoded in them.
This qubit scrambles information across the system, but initial information is 
preserved in the wave function in the SG-MBL phase.
How robustly this picture of the bulk qubit holds in the unitary time evolution
by the Hamiltonian, $\HIC$, is an interesting problem.
On the other hand for random circuit of stabilizers, we think that the bulk qubit is a good picture during time evolution.

Here, we emphasize that the Pauli spins at the edges of the 
open boundary chain with the length $L$, $\s^x_{1(L)}$, commute with the non-interacting part of the SG-MBL Hamiltonian with $h_i=0$, 
$\sum_iJ_{i}\s^x_i\s^x_{i+1}$, and anti-commute with $\mathbb{P}$.
Then, $\s^x_{1(L)}$ is a zero mode operator from the viewpoint of topological 
order \cite{Fendley2012}.
The operation of $\s^x_{1(L)}$ on the above two cat states interchanges them, respecting 
${\bf Z}_2$ parity symmetry $\mathbb{P}$.
Even for finite $\{h_i\}$, the zero-mode operator can be constructed perturbatively \cite{Fendley2012}, such as 
$\s^x_1+{h_1 \over J_{1}}\s^z_1\s^x_2+{h_1h_2 \over J_{1}J_{2}}\s^z_1\s^z_2\s^x_3+\cdots$.
[In the Majorana representation, gapless edge mode $\gamma_1 = \s^x_1$.] 
Further, this zero mode can survive even in the presence of a finite interaction $g$,
and its explicit form is obtained perturbatively such as\cite{Kemp2017},
$\s^x_1+{h_1 \over J_1}\s^z_1\s^x_2+{g\over J_2}\s^y_1\s^y_2\s^x_3 +\cdots$.

In order to verify the property of the phases furthermore,
we explore the spin-glass order by studying the spin correlation, $G_{r=L/2}$,
defined by \cite{Gergs2016} 
\be
G_r = {1 \over L-r} \sum^{L-r}_{i=1} |\s^x_i\s^x_{i+r}|.
\label{Gr}
\ee
The result shown in Fig.~\ref{Fig1}(b) indicates that the spin-glass order emerges 
as $\d$ increases from $\d_c$, as we expect.

\begin{figure}[t]
\begin{center} 
\includegraphics[width=7.0cm]{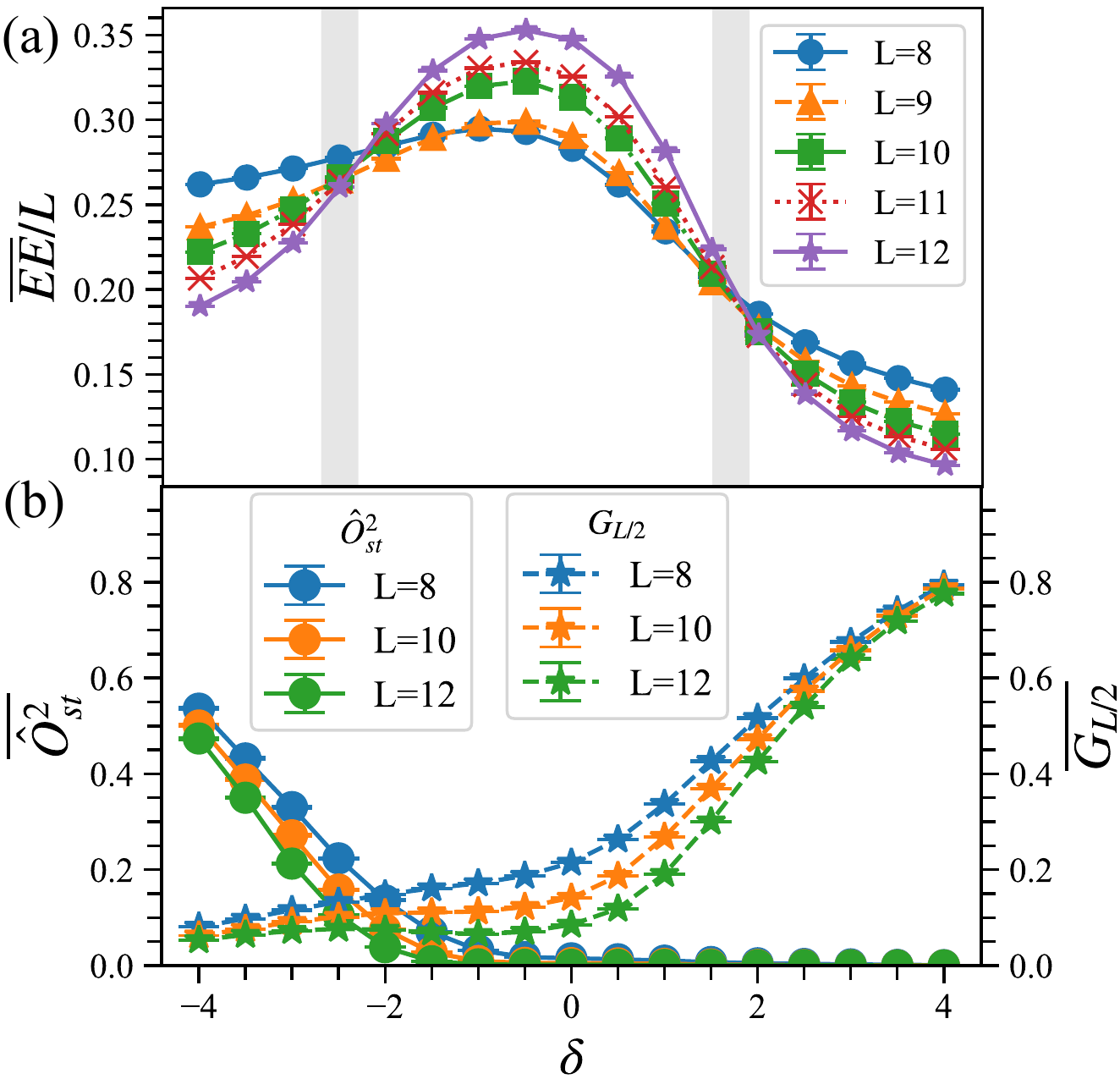}  
\end{center} 
\caption{(a) Half-chain EE for the cluster spin chain model $\HCS$ in Eq.~(\ref{HCS})
for $g = 0.2$.
Calculation of $\overline{EE}/L$ indicates the existence of two phase transitions
such as the CS-MBL $\to$ ETH $\to$ SG-MBL as $\d$ increases.
As $\d$ is getting large, $\overline{EE} \to \log 2$ corresponding to the SG-order,
whereas as $\d$ decreases, $\overline{EE} \to \log 4$ coming from `emergent'
$\mathbb{Z}_2\times \mathbb{Z}_2$ symmetry. 
These results were obtained by averaging over $1000, 750, 500, 300,$ and $150$ disorder realization using all eigenstates for the $L = 8, 9, 10, 11, 12$ systems. 
(b) Spin correlation, $G_{L/2}$ in Eq.~(\ref{Gr}), as a function of $\d$. 
Its increase indicates the SG-order for $\d \gg1$.
On the other hand, the string order, $\Phi_{\rm st}$ acquires non-vanishing values
for $\d<-2$ indicating topological order with $\mathbb{Z}_2\times \mathbb{Z}_2$ symmetry.
These results were obtained by averaging over the 20000 eigenstates using 10-20 eigenstates in the middle of the spectrum at each disorder realization for the L=8, 10, and 12 systems.}
\label{Fig2}
\end{figure}

The second spin chain system, which we call extended random cluster spin 
chain \cite{Bahri}, is described by the following Hamiltonian,
\be 
H_{\rm CS} &=& \sum_i \Big[ J_i \s^x_i \s^x_{i+1} + \l_i \s^x_{i-1} \s^z_i \s^x_{i+1}+ \tilde{h}_i \s^z_i\Big]  \nonumber \\
&& + H_g,   
\label{HCS}
\ee
where $\tilde{h}_i$'s are small random variables drawn from $[0, 1]$, and $\l_i$ are uniform random variables
drawn from $[0, W_\l]$.
We define $W_J =(W_\l)^{-1}=W$ and also $\delta = 2 \ln W$ as before.
As we showed in the above, for sufficiently large $W_J$, the all states in $\HCS$ belong 
to the SG-MBL.
On the other hand for sufficiently large $W_\l$, $\HCS$ approaches the random cluster 
spin model,
which is a symmetry-protected topological (SPT) system with the 
$\mathbb{Z}_2\times \mathbb{Z}_2$ symmetry \cite{Tasaki2021}. 
Also, a similar disordered model has been studied
and clarified its ground state phase diagram in terms of the disorder-strength parameter
space \cite{Lieu2018}, where the SPT phase is characterized by the number of the zero
energy Majorana edge modes.
It is also known that energy eigenstates of the genuine cluster spin model with only 
second terms of Eq.~(\ref{HCS}) are all localized as dictated by 
LIOMs, $\{\s^x_{i-1} \s^z_i \s^x_{i+1}\}$, and we shall verify in the subsequent
calculation that this localization nature remains for small but finite values of
$\{J_i\}$.
Here, we again emphasize that
the above LIOMs are nothing but stabilizers in quantum information theory \cite{Briegel_2001}.
Since the single stabilizer takes two eigenvalues $\pm 1$, the operator 
can be regarded as a logical spin operators, that is, a qubit.
In what follows, we call them stabilizer-qubits . 
In the random Ising spin chain with $W\gg 1$, the stabilizer-qubits are $\{\sigma^{x}_i\sigma^{x}_{i+1}\}$.
The stabilizer-qubit is one of key concepts for understanding findings in the present 
work as we explain. In contrast to general LIOMs in the conventional MBL, stabilizer qubit realizes localization with some order, e.g., SG or topological order.

To obtain the phase diagram of the system $\HCS$ in Eq.~(\ref{HCS}), we first investigate 
the half-chain EE,
and display the numerical calculations in Fig.~\ref{Fig2}(a).
The results show that there are three phases, i.e., two MBL phases and one thermal phase.
Interestingly enough, for $\delta \ll -1$, the EE approaches $\log 4$ instead of $\log 2$,
although
in the presence of the $J_i$-terms as well as the $g$-terms, the Hamiltonian $\HCS$ 
has only $\mathbb{Z}_2$ symmetry.
In order to verify the topological properties of the phase, we calculate a string order, defined as 
${\cal O}_{\rm st}(i,j) \equiv \langle \s^x_i\s^y_{i+1}(\prod^{j-2}_{k=i+2}\s^z_k)\s^y_{j-1}\s^x_j\rangle$.
The results of the string order averaged over the randomness, 
$\Phi_{\rm st} \equiv \overline{{\cal O}_{\rm st}^2}$,
are displayed in Fig.~\ref{Fig2}(b), which indicate that the topological order
corresponding to
the genuine cluster spin model exists for $\delta \ll -1$. 
This is an unexpected result since the finite $J_i$-terms reduce the symmetry from
$\mathbb{Z}_2\times \mathbb{Z}_2$ to $\mathbb{Z}_2$.
However, a similar result is observed for the ground state in Ref.~\onlinecite{Smacchia}.
In the clean system, $\Phi_{\rm st}$ for the excited state vanishes as expected 
from the observation in Ref.~\onlinecite{Huse2013}.

In Fig.~\ref{Fig2}(b), 
we also show the calculations of the spin-glass order parameter, $G_{L/2}$.
We find a similar behavior to that in the random Ising spin chain in Fig.~\ref{Fig1} (b).
That is, 
finite value of $G_{L/2}$ for $\d>2$ indicates the existence of the spin-glass order, 
and this result is obviously in good agreement with the EE in Fig.~\ref{Fig2}(a).

We investigated the infinite-temperature phase diagram of the $g=0$ system
of $H_{\rm CS}$ in Eq.~(\ref{HCS}).
Similarly to the random Ising spin chain, the calculation of the EE seems to indicate
the direct transition from the SG-MBL to the cluster-spin (CS)-MBL. 
We recently investigated a very close model to  $H_{\rm CS}$ in Eq.~(\ref{HCS})
by using a Majorana fermion~\cite{Kuno2022NJP}.
Similar methods can be applied to the model $H_{\rm CS}$ with $g=0$.
[Please see later discussion in Sec.~III.D.]
The model reduces to a random-hopping and random potential free Majorana fermion, which
is similar to the TRIM case~\cite{Fisher1995} and is expected to  exhibit 
a phase transition via varying the strength of the random hopping and potential.

\subsection{Tripartite mutual information}
\begin{figure}[t]
\begin{center} 
\includegraphics[width=6.5cm]{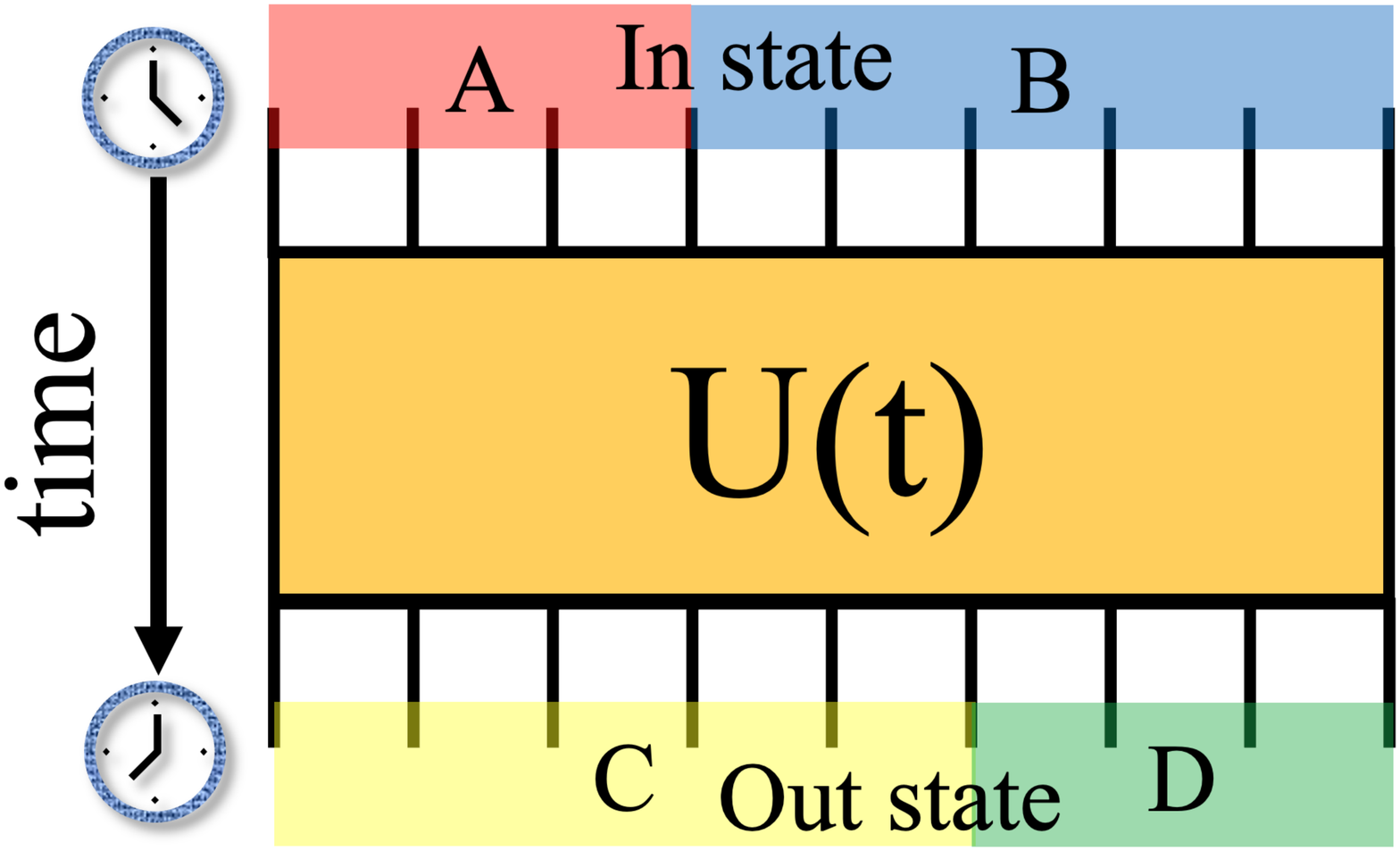}  
\end{center} 
\caption{Schematic image of the time evolution of the state with doubled Hilbert space. 
The spatial partitioning of the system is represented where four subsystems A, B, C, and D are introduced.}
\label{Fig3}
\end{figure}

In the previous section, the half-chain EE, spin-glass and string orders identified 
the MBL phases and thermal phase in the models. 
Next, we investigate the property of information spreading in each phase.
We observe how the information spreading takes place in each phase and how strongly
the time-evolution operator works as a scrambler. 
To quantify the information spreading ability in the system, 
we employ tripartite mutual information (TMI), which is very efficient tool to evaluate the scramble ability of the unitary time evolution operator of the system, as proposed in Ref.~\onlinecite{Hosur}. 
There are already some observations of the TMI in an interacting model and conventional
MBL systems \cite{Schnaack2019,Mascot2020,MacCormack2021,Bolter2022,KOI2022}. 
In addition, the observation of the TMI can be an efficient indicator to characterize a phase transition (phase boundary) in the context of the measurement induced phase
transition \cite{Zabalo2020}.

Let us explain the TMI and the practical methods of the numerical calculation \cite{KOI2022} 
to be applied for the target models with $L$ lattice sites. 
Our numerical resource allows us to calculate the TMI up to the system size $L=12$ by the methods.

By using the TMI, we can quantify the information spreading and scrambling embedded in the time evolution operator 
$\hat{U}(t)\equiv e^{-itH}$, where $H$ is either $\HIC$ or $\HCS$ in this work.
On calculating the TMI, we use the state-channel map that plays an essential role. 
Under this map, the operator $\hat{U}(t)\equiv e^{-itH}$ 
is regarded as a pure quantum state in the doubled Hilbert space,
${\cal H}_{\rm D} \equiv{\cal H}_{\rm in} \otimes {\cal H}_{\rm out}$~\cite{Hosur}.
We start from the density matrix at time $t$, 
$\rho(t)=\sum^{N_D}_{\nu=1}p_{\nu}\hat{U}(t)|\nu\rangle \langle\nu|({\hat{U}(t)})^{\dagger}$, 
where $\{|\nu\rangle\}$ is a set of a orthogonal bases state (time independent),
$N_D$ is the dimension of the Hilbert space in the system, and an 
input ensemble is encoded by parameters $\{ p_\nu\}$. 
Then, by applying the state-channel map to the density matrix $\rho(t)$,
the time-evolution operator is mapped into a pure state in the doubled Hilbert space,
\begin{eqnarray}
\rho(t)\to 
|U(t)\rangle=\sum_{\mu}\sqrt{p_\nu}(\hat{I}\otimes {\hat U}(t))
|\nu\rangle_{\rm in}\otimes|\nu\rangle_{\rm out},
\label{pure_state_U}
\end{eqnarray}
where $\hat{I}$ is the identity operator and $\{|\nu\rangle_{\rm in}\}$ and
$\{|\nu\rangle_{\rm out}\}$ are the same set of orthogonal bases state.
The time evolution operator ${\hat U}(t)$ acts only on the out orthogonal states $|\nu\rangle_{\rm out}$.
An arbitrary input ensemble can be employed by tuning $\{ p_\nu\}$ \cite{Hosur}.
In this work, however, we mostly focus on the infinite temperature ensemble, i.e.,
${p_{\nu}}=1/N_{D}$ to see universal properties of the time-evolution unitary. 
Then at $t=0$, as $\hat{U}(0)=\hat{I}$, the in-state and out-state are maximally entangled.
To calculate the TMI under the time evolution with the Hamiltonian $\HIC/\HCS$, 
spatial partitioning of the pure state $|U(t)\rangle$ has to be specified. 
The spatial partitioning is done for both the $t=0$ in-state and the out-state at $t$
respectively. 
Figure~\ref{Fig3} shows that the $t=0$ state (given by $\rho(t=0)$) is divided into two
subsystems $A$ and $B$, 
and the state at time $t$ (given by $\rho(t)$) is divided into two subsystems $C$ and $D$.
In later calculations, we mostly focus on the partition with the equal length of $A$ and
$B$ (and also $C$ and $D$) subsystems, as well as asymmetric one for specific purposes.
(See later discussion.)

Under this spatial partitioning, the density matrix of the pure state 
$|U(t)\rangle \in {\cal H}_{\rm D}$ 
is denoted as $\rho_{ABCD}(t)=|U(t)\rangle \langle U(t)|$.
From this full density matrix $\rho_{ABCD}(t)$, 
a reduced density matrix for a subsystem $X$ is obtained by tracing out the degrees of
freedom in the complementary subsystem of $X$ denoted by ${\bar X}$, i.e.,
$\rho_{X}(t)=\mathrm{tr}_{\bar X}\rho_{ABCD}$. 
From the reduced density matrix $\rho_{X}(t)$, the operator entanglement entropy (OEE) for 
the subsystem $X$ is obtained by conventional von-Neumann EE, 
$S_X=-\mathrm{tr}[\rho_{X}\log \rho_{X}]$.
From the OEE, we introduce the bipartite mutual information (BMI) of $X$ and $Y$ 
subsystems 
(where $X, Y$ are some elements of the set of the subsystems $\{A,B,C,D\}$, 
and $X\neq Y$); 
\begin{eqnarray}
I(X:Y)=S_X+S_Y-S_{XY}.
\label{MI}
\end{eqnarray}
The value of $I(X:Y)$ quantifies how the subsystems $X$ and $Y$ correlate with each other.

By using the BMI, the TMI for the subsystems $A$, $C$ and $D$ is 
defined as; 
\begin{eqnarray}
I_3(A:C:D)=I(A:C)+I(A:D)-I(A:CD).
\label{TMI_def}
\end{eqnarray}
The above TMI quantifies how the initial information embedded in 
the subsystem $A$ spreads into both subsystems $C$ and $D$ in the output state. 
If the spread of the information in $A$ sufficiently occurs across the entire system
at time $t$, $I_3(t)$ gets a negative value, while 
the BMI keeps a non-negative value even in such a situation.
In general $I_3$ is zero at $t=0$, as $|U(0)\rangle$ is the product state of the EPR pair
at each lattice site.
When the time-evolution operator acts as a strong scrambler, $I_3$ acquires a large
negative value under the time evolution. 
On the other hand, if the time evolution operator
does not act as an efficient scrambler, $I_3$ remains small.  
Hence, $I_3$ is a good indicator to quantify the degree of scrambling, i.e., the information spreading. 
In this paper, we mostly employ the TMI to characterize the scrambling for our target
models, as well as quench dynamics of the EE.

In the following numerical calculations, 
it is convenient to set a reference frame of the TMI, $I_3$, as in Refs.~\onlinecite{Schnaack2019,Bolter2022}. 
The reference flame is the value of the TMI of the Haar random unitary, $I^{H}_{3}$, 
which depends on the Hilbert space dimension of the system size $L$ \cite{Haar_ND}. 
The value of $I^{H}_{3}$ can be numerically calculated \cite{Haar_val}. 
Then, we define a normalized TMI, $\tilde{I}_3(A:C:D)$, as follows,
\begin{eqnarray}
\tilde{I}_3(A:C:D)\equiv \frac{I_3(t)}{I^{H}_{3}}.
\label{TMI_def}
\end{eqnarray}
In the following sections, we numerically obtain the value of $\tilde{I}_3$.

Here, some remark is in order.
In the practical calculation, we do not directly obtain the density matrix in the doubled
Hilbert space, $\rho_{ABCD}(t)$. 
Instead, some specific methods are utilized in order to study systems as large as  
possible by our numerical resource. Details are explained in 
our previous paper~\cite{KOI2022}. 
In the following numerical calculations, we also employ the Quspin solver \cite{Quspin}
to efficiently construct the numerical basis and time evolution operators.

\section{Numerical studies}

In this section, we shall perform the systematic numerical study by observing the quench dynamics of the EE and the information spreading quantified by the TMI. 
We show typical dynamical aspects inherent in both systems, $H_{\rm IC}$ and $H_{\rm CS}$.
The numerical investigation of the models uncovers initial state dependence of
the quench dynamics of the EE, which is strongly related with duality in the random Ising
spin chain, and also it clarifies characteristic behavior of the TMI for systems with
topological order. 
In particular, the calculation of the TMI is independent of the choice of initial state and exploits essential properties of the scrambling embedded in the unitary time-evolution operator: 
(I) We capture distinct phase transitions and their phase boundary. 
(II) By varying the size of the partitioning in the calculation of the TMI, 
we can extract the bulk structure of information spreading for both topological MBLs,
corresponding to the degree of the scrambling. 
The SG-MBL and CS-MBL phases can be clearly distinguished from this aspect. 
In what follows, we set a unit of time $\hbar/W$ in numerical calculations of quench dynamics.

\subsection{Quench dynamics of bipartite EE: random Ising spin chain}

We start to show the numerical results of the quench biparite EE of the system 
$H_{\rm IC}$ at infinite temperature for $g=0$ and $g=0.2$.
The case of $g=0$ is the TRIM, and the IRC point at
$\d=0$ separates the paramagnetic and spin glass localized phases \cite{Fisher1995}.  
The ground state for an arbitrary $\d$ is the Griffiths state in which both typical 
length and typical time scale have very broad distributions \cite{Fisher1995,Young1996,Fisher1999}.
This gapless Griffiths phase persists at finite temperature, as well as the spin-glass
order for $\d \gg 1$.
Therefore, it is interesting to see how entanglement spreads in that specific regime.
For the case of the interacting case with $g>0$, on the other hand,
the ergodic state intervenes between the two MBL states, which are connected by duality.
How the entanglement entropy spreads in the states connected by duality is an interesting
problem, and it sheds light on quantum information spreading as we see later on.
 
\begin{figure}[t]
\begin{center} 
\includegraphics[width=7.0cm]{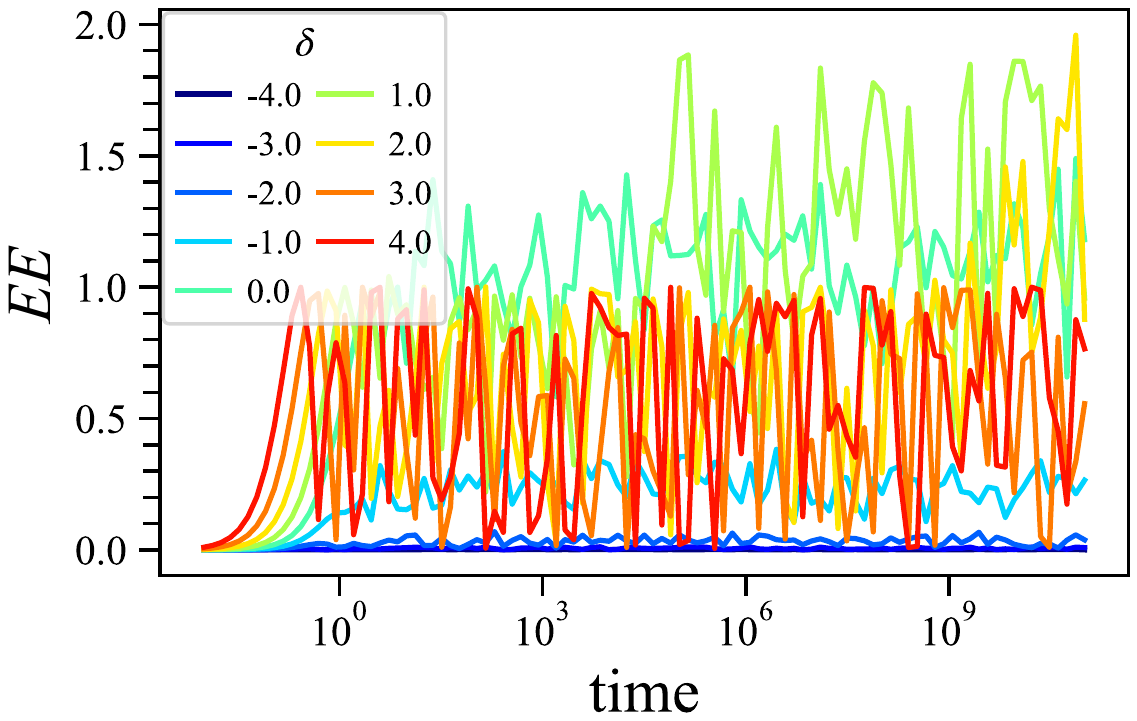}  
\end{center} 
\caption{Quench dynamics of the entanglement entropy, $\Et$: 
the random Ising spin chain, $H_{\rm IC}|_{g=0}$ in Eq.~(\ref{HIC}).
For $\d\gg 1$, $\Et$ oscillates quite rapidly, whereas for $\d\ll -1$,
it keeps very small values.
The system with $\d=0$ corresponds to infinite randomness critical point.
The system size is $L=12$.
}
\label{Fig4}
\end{figure}

We study the quench dynamics in this subsection, i.e., the time evolution of 
the half-chain von Neumann entropy,
$\Et$ obtained from a time evolved state.
We first consider the non-interacting case of the random Ising chain with $g=0$, 
the TRIM.
The quench half-chain EE, $\Et$, is defined as follows;
\be
\Et = - \mbox{Tr} [\rho_{\rm r}(t)\log (\rho_{\rm r} (t))],
\label{Sent}
\ee
where $\rho_{\rm r} (t)$ is the reduced density matrix of the half chain at time $t$.
Let us investigate the case in which the employed initial state is
$|\ua\ua\ua\cdots\rangle_Z$ in the $\s^z$-basis. 
The results in Fig.~\ref{Fig4} show that $\Et$ for $\d=-4.0$ and $-3.0$ keeps 
a very small value during the time evolution, and 
$\Et$ for the other $\d$'s exhibits rather strong oscillating behavior.
The averaged values of $\Et$ in the central regime of $\d$ are larger than those of
$\d=4.0$ and $3.0$.
This \textit{dynamical behavior} obviously reflects the IRC at $\d=0$.
We observed similar behavior of $\Et$ for the initial state $|\ua\da\ua\da\cdots\rangle_Z$
(not shown).
This strong oscillation of $\Et$ is an unusual one and is expected to reflect the
Griffiths properties of the states. 
For the case of $\d=-4.0, -3.0$, the random field dominates the bond coupling, and
therefore,
a phenomenon similar to Anderson localization takes place there with vanishingly small
$\Et$.

\begin{figure}[t]
\begin{center} 
\includegraphics[width=7.0cm]{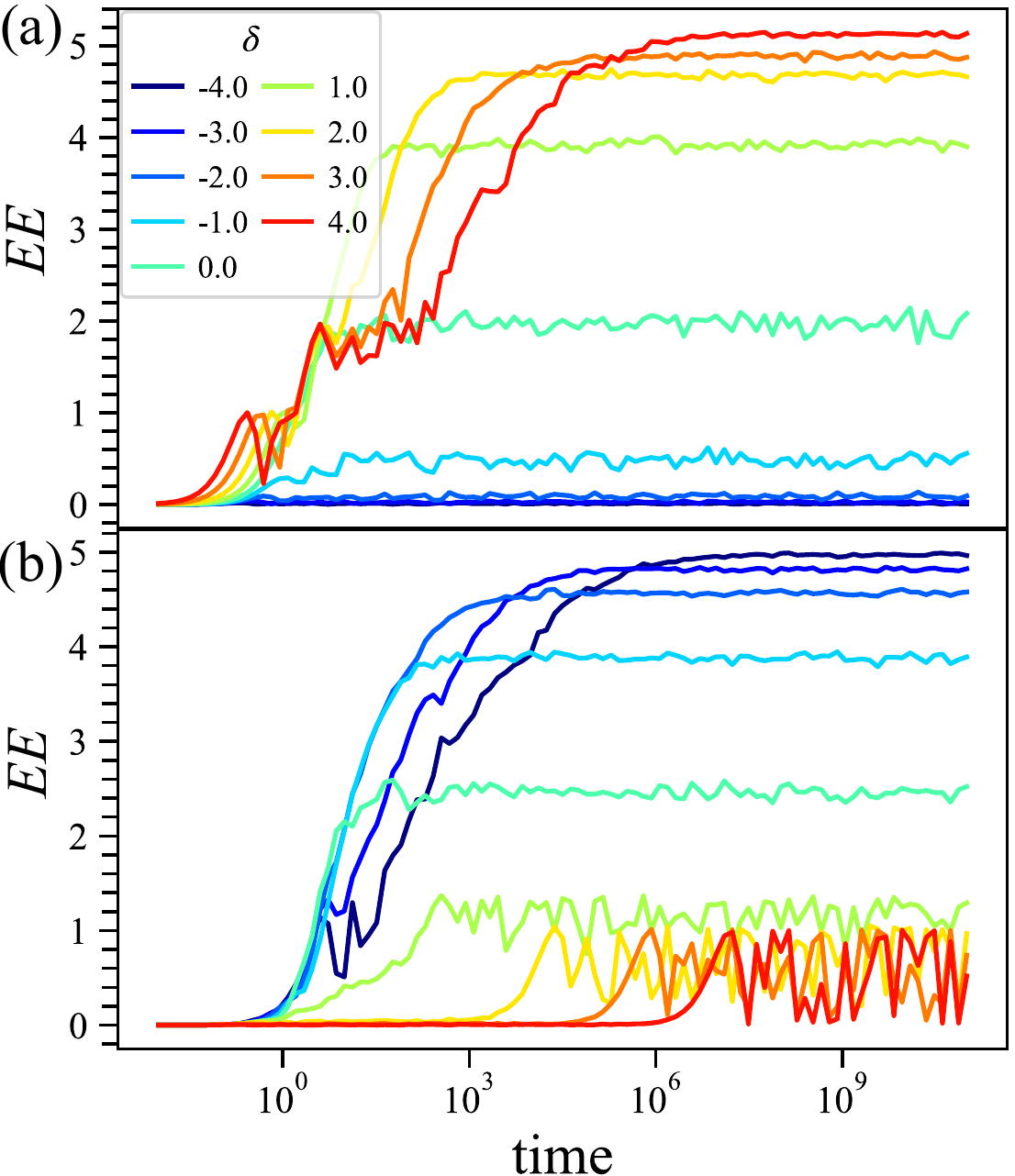}  
\end{center} 
\caption{Quench dynamics of the entanglement entropy, $\Et$: 
the random Ising spin chain with $g=0.2$ for $L=12$.
(a) initial state: $|\psi(t=0)\rangle=|\ua\ua\ua\cdots\rangle_Z$,
(b) initial state: $|\psi(t=0)\rangle=|\ua\ua\ua\cdots\rangle_X
=\prod_i\frac{1}{\sqrt{2}}(|\ua\rangle_i+|\da\rangle_i)_Z$.
}
\label{Fig5}
\end{figure}

Let us turn to the interacting case with $g=0.2$.
In Figs.~\ref{Fig5}(a) and (b), we display the calculations of $\Et$ for the initial
states $|\ua\ua\ua\cdots\rangle_Z$
in the $\s^z$-basis and also $|\ua\ua\ua\cdots\rangle_X$ in the $\s^x$-basis, respectively.
We first note that the state for $g=0.2$ does not have the Griffiths-state nature, 
as $\Et$ is quite stable compared with the non-interacting case.
Figure \ref{Fig5} shows interesting behaviors of $\Et$, that is, for the initial 
state $|\ua\ua\ua\cdots\rangle_Z$, 
$\Et$ for $\d=4.0, 3.0$ acquires large values in the time evolution, whereas for 
$|\ua\ua\ua\cdots\rangle_X$, $\Et$ for $\d=-4.0, -3.0$ increases similarly and 
saturates to large values.
This result indicates that the bond coupling, $\sum_i\s^x_i \s^x_{i+1}$, dominates 
the field coupling,
$\sum_ih_i\s^z_i$, for $\d\gg 1$, and the states $|\ua\ua\ua\cdots\rangle_Z$ contains 
all states of the $\s^x$-basis,
then as a result, the EE is generated in the time evolution.
The same thing happens for the case with $\d \ll -1$ and $|\ua\ua\ua\cdots\rangle_X$, 
in which the field coupling dominates the bond coupling.
We can understand the above behavior of $\Et$ from duality.
In the Hamiltonian level, the random parameters $\{J_i\}$ and $\{h_i\}$ are interchanged 
by Eq.~(\ref{Dutr}).
The above numerical study of $\Et$ shows that duality transformation of the initial 
state is needed for $\Et$ to exhibit similar behavior in the corresponding duality
counterparts.

A careful look at Fig.~\ref{Fig5} (b) reveals some important aspect of the time evolution 
of $\Et$, besides the above increasing behavior.
That is, $\Et$ for $\d=4.0, 3.0$ and the initial state $|\ua\ua\ua\cdots\rangle_X$ 
has small but finite values for the late time of the time evolution.
On the other hand,  $\Et$ for $\d=-4.0, -3.0$ and the initial state
$|\ua\ua\ua\cdots\rangle_Z$ in Fig.~\ref{Fig5} (a)
keeps vanishingly small values in the time evolution.
This result seems to break duality of the Hamiltonian $\HIC$.
We think that this discrepancy comes from the topological order of 
the SG-MBL, which is
observed through the EE in Sec.~II, i.e., 
the topological order exhibits a long-range correlations \cite{EE_long_range}
characterized by non-local order parameter, such as string order,
and it possibly enhances information spreading across the almost entire system compatibly
with MBL identified by the return probability, etc. 
More explicitly in the SG-MBL phase, such a non-local order may be construct. 
That is, we can consider a string
operator, given by $\langle\prod_{k=i}^{j-1}\s^x_k\s^x_{k+1}+\cdots\rangle
=\langle\s^x_i\s^x_j\rangle+\cdots$, as in the cluster spin chain discussed in Sec.~II. 
Please note that the leading terms of the LIOMs are given by the dimer,
$\{\s^x_i\s^x_{i+1}\}$, in the SG-MBL regime. 
The above observation clarifies the relationship between the spin-glass order and
the topological order, i.e., the topological order accompanies the long-range spin-glass
correlation described by $G_r$.
The long-range correlation makes a pair of large qubit by $\mathbb{Z}_2$ symmetry, 
and their mixing emerges in late-time evolution as seen in Fig.~\ref{Fig5}(b).

In order to understand the above observation for the EE of the SG-MBL state more
concretely, let us consider
a four-spin system and divide it into two two-spin subsystems, i.e., $A$ and $B$ subsystems.
Then, the initial state corresponding to Fig.~\ref{Fig5}(b) is given by,
\be
|\psi_0\rangle &=&|\ua\ua\ua\ua\rangle_X  \nonumber \\
&=&{1 \over 2}[\psi(A,+)+\psi(A,-)]  [\psi(B,+)+\psi(B,-)], \nonumber\\
\label{psi0}
\ee
where $\psi(A,\pm)={1\over \sqrt{2}}(|\ua\ua\rangle_A\pm |\da\da\rangle_A)$
[energy eigenstates of $A$ subsystem expressed in the $X$-basis], 
and similarly for $\psi(B,\pm)$.
In the time evolution, other states such as $|\ua\da\rangle_A$ emerge only as a
perturbation (by $\{h_i\s^z_i\}$) because of the existence of the stabilizer, whose
leading terms are given by $\{\s^x_i\s^x_{i+1}\}$ ($|J_i| \gg |h_i|, g$).
Then, by ignoring perturbative states, the system can be regarded as a system 
of two quantum degrees of freedom with two quantum states for each.
Entanglement entropy of this kind of system was studied in Ref.~\onlinecite{Serbyn2013}.
The interactions between $A$ and $B$ subsystems are given by $J_2\s^x_2\s^x_3$ and
$g(\s^x_1\s^x_3+\s^x_2\s^x_4+\s^z_2\s^z_3)$.
These interactions are invariant under $\mathbb{P}$. 
Especially, $\s^x_2\s^x_3$, $\s^x_1\s^x_3$ and $\s^x_2\s^x_4$ terms generate
mixing between
the states $\psi(A,+)\psi(B,+)\leftrightarrow\psi(A,-)\psi(B,-)$ and 
$\psi(A,+)\psi(B,-)\leftrightarrow\psi(A,-)\psi(B,+)$
as they operate such as
$|\ua\ua\rangle_A|\da\da\rangle_B \to - |\ua\ua\rangle_A|\da\da\rangle_B$, etc.
[The term $\s^z_2\s^z_3$ works only as a perturbation as $\{h_i\s^z_i\}$.]
This mixing obviously generates an extra time dependence in each of the four terms
in Eq.~(\ref{psi0}) [$\psi(A,+)\psi(B,+), \cdots, \psi(A,-)\psi(B,-)$]
and a non-trivial reduced density matrix, 
and as a result, the oscillating EE emerges~\cite{Serbyn2013}.
In the original many-body system, the wave functions corresponding to $\psi(A,\pm)$, 
etc., 
have a complicated form
under the time evolution, and the reduced density matrix $\rho_A$ is of high dimensions.
However, we expect that an oscillating behavior originating from the above mechanism persists. 

Returning to the SPT order, we note that for $i=1$ and $j=L$, the string operator
essentially measures
the correlation between edge operators, $\langle \s^x_1\s^x_L\rangle$, mentioned
in Sec.~II. 
This expression of the string order is obviously a reminiscence of the Stokes' theorem
by which a magnetic flux piercing a surface is expressed by a line integral of vector potential along the boundary.

In addition, we should remark that duality is explicitly broken at edges in the
open-boundary system,
and therefore, duality does \textit{not} respect the relation between the bulk and its
edges.

In the following subsection, we shall study the TMI. 
The above observations will shed light on the results of the TMI.

\begin{figure}[t]
\begin{center} 
\includegraphics[width=7.0cm]{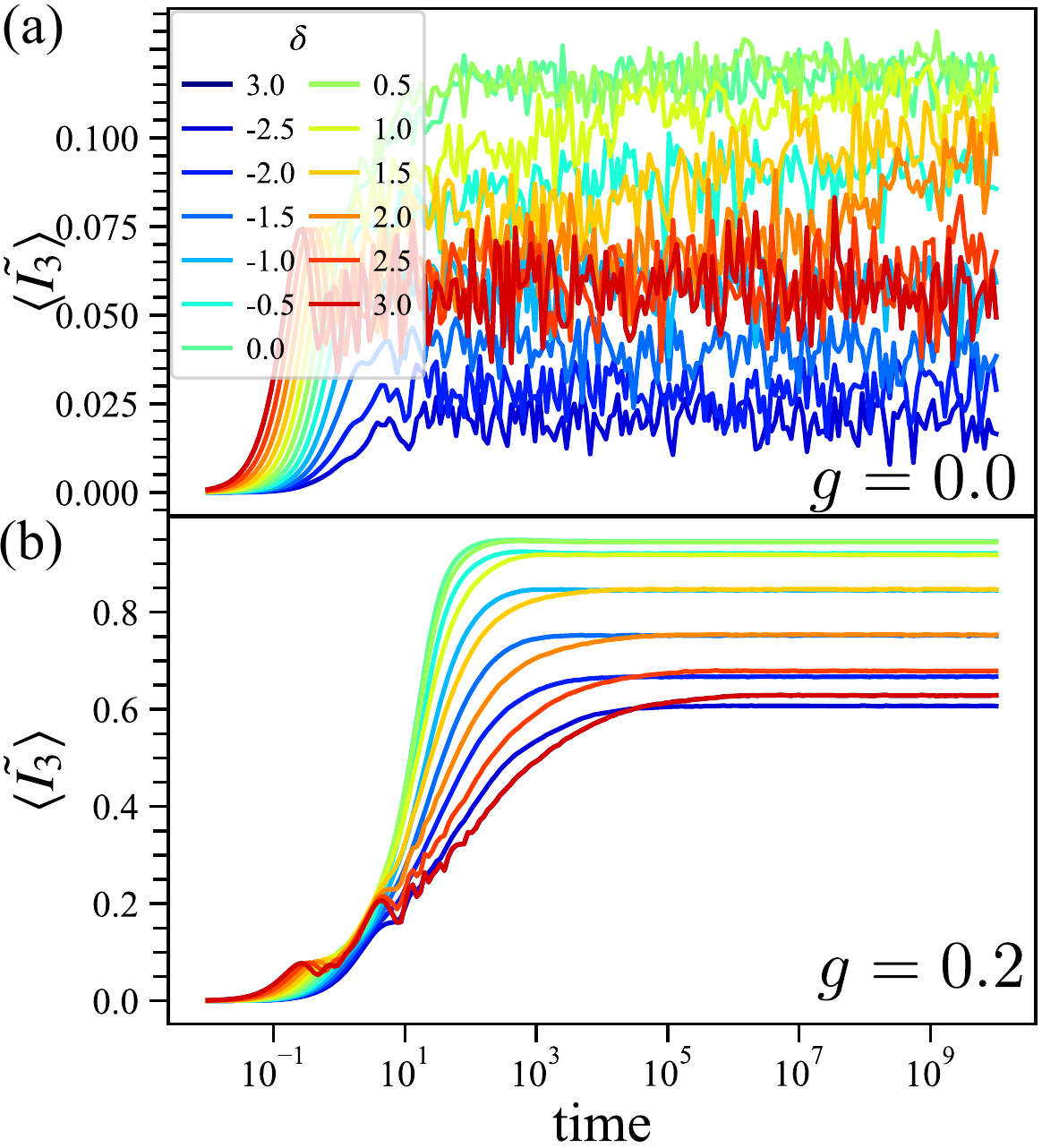}  
\end{center} 
\caption{TMI dynamics of the random Ising spin chain, $\HIC$.
(a) The non-interacting case with $g=0$. 
The TMI exhibits oscillating behavior for all $\d$'s, but its short-period time
average is a stable function of time.
(b) The interacting case with $g=0.2$.
After early-time increases, the TMI saturates to a finite value for each value of $\d$. 
In both cases, the TMI has larger values in the regime $\d\sim 0$ compared to
other regimes. These results were obtained by averaging over the $10$ disorder realization for $L=12$ system. Here, $\langle\cdots\rangle$ denotes disorder average.}
\label{Fig6}
\end{figure}

\subsection{Tripartite mutual information: Ising spin chain}

In the previous subsection, we studied the bipartite EE, $\Et$, for the non-interacting
($g=0$) as well as the interacting case ($g=0.2$), and obtained interesting results, 
in particular, from the viewpoint of duality and SPT order.
In this subsection, we shall study the behavior of the TMI under the time evolution.
In the practical calculation, the system size is $L=8, 10$ and $12$, and the spin chain 
is divided into two chains with equal length to compute $I_3(A:B:C)$. 
In the numerical calculation, we focus on the disorder average of the normalized TMI $\tilde{I}_3$ 
denoted by $\langle \tilde{I}_3\rangle$.

\begin{figure}[t]
\begin{center} 
\includegraphics[width=7.0cm]{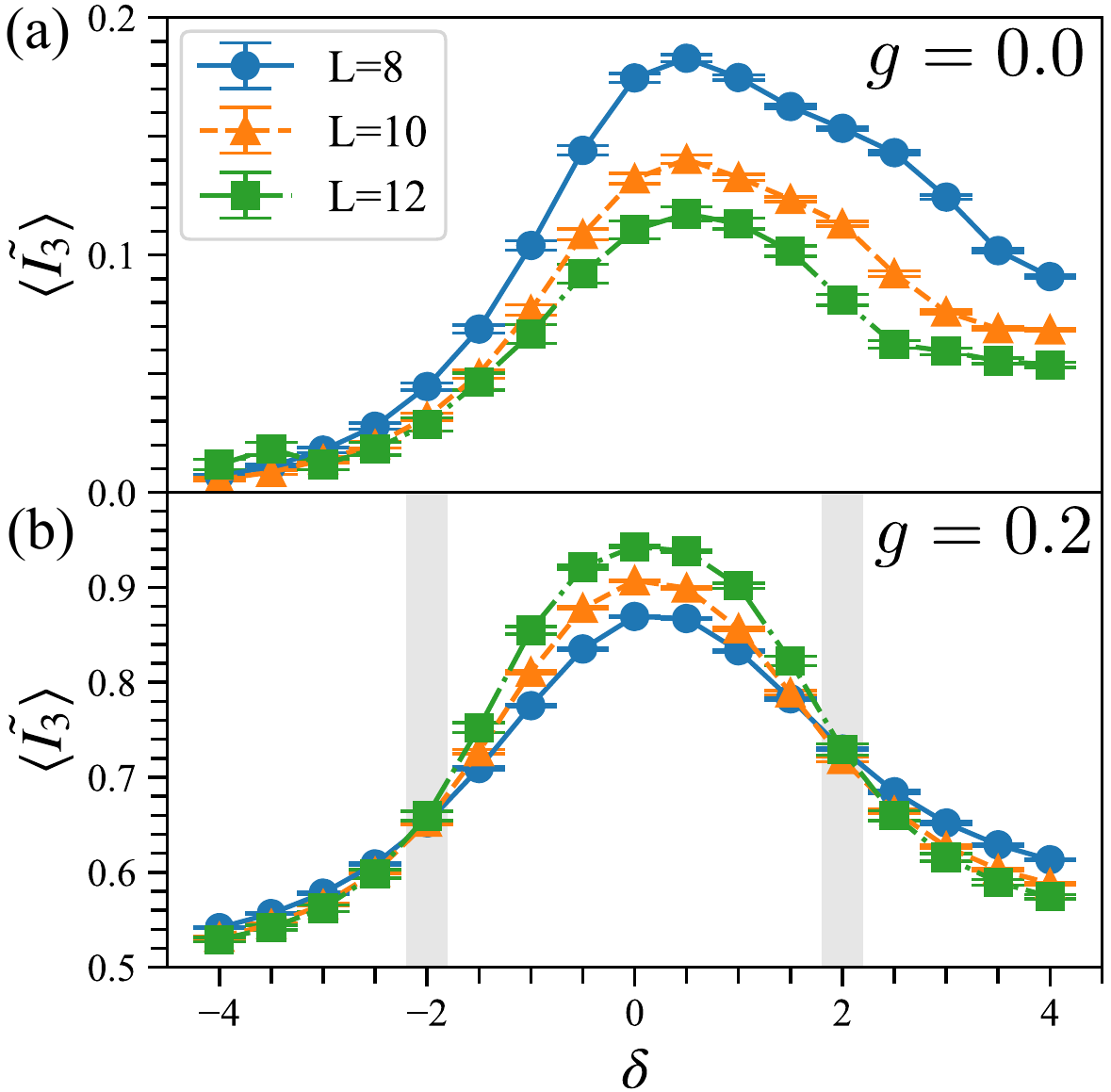}  
\end{center} 
\caption{Saturation values of TMI for various system size of random Ising spin chain. 
(a) $g=0$ case: 
$\tilde{I}_3$ exhibits a peak at $\d=0$, which corresponds to the infinite
random critical point of the random transverse Ising chain.
Curves of $\tilde{I}_3$ do not cross with each other indicating non-existence
of phase transitions besides $\d=0$.
(b) $g=0.2$ case: 
$\tilde{I}_3$ for the interacting case. 
Curves of $\tilde{I}_3$ cross with each other at the phase transition points,
$\d_{c1} \simeq -\d_{c2}\simeq -2.0$.
Duality symmetry is obviously broken in the MBL regimes.
These results were obtained by averaging over the $1000, 500,$ and $100$ disorder
realization for L=8, 10, 
and 12 systems. We define saturation values of TMI as the average of $I_3(t)$ 
at $10$ points between $t=10^9$ and $t=10^{10}$.
}
\label{Fig7}
\end{figure}

We study the behavior of the TMI for fixed values of $g$ by varying $\d$ to see
how it behaves in the various phases.
As we explained in Sec.~II, we consider the infinite-temperature ensemble.
In Figs.~\ref{Fig6}, we show the time evolution of the TMI, $\tilde{I}_3$, for 
the non-interacting and interacting 
Ising spin chains [$\HIC$ in Eq.~(\ref{HIC})] with various values of $\d$.
The two cases exhibit quite different behavior, the strong oscillation in the
non-interacting, and 
stable behavior in the interacting case, although the time average of both of them is 
rather stable and is an increasing function of time. 
The strong oscillation of $\tilde{I}_3$ in the non-interacting case comes from 
the Griffiths nature of the broad 
distribution of the localization length and typical time scale as the above calculation 
of $\Et$ shows.
However, $\tilde{I}_3$ has stable values in short-period time average, 
which depends on the parameter $\d$ [not shown].
Careful look at the calculations in Fig.~\ref{Fig6}(a) reveals that $\tilde{I}_3$ 
increases even after $t=10^9$, in particular, $\d=2.0$ and $2.5$.
See further late-time calculation in Appendix, where we show the late-time behavior of $\tilde{I}_3$ for $\d=2.0$ and $2.5$,
and find instability of $\tilde{I}_3$, that is, which does not saturate.
However, its system-size dependence is quite stable, and we think that this
observation guarantees reliability of the result shown in Fig.~\ref{Fig7}(a).
On the other hand in the interacting case of $g=0.2$, $\tilde{I}_3$ is an increase
function and saturate into stable values depending on $\d$ after the early-time evolution.

In Figs.~\ref{Fig7}(a) and ~\ref{Fig7}(b), 
we show the saturation values of $\tilde{I}_3$ 
as a function of $\d$ and also
exhibit its system-size dependence for both non-interacting ($g=0$) and interacting
cases ($g=0.2$).
For both cases, $\tilde{I}_3$ has a peak at $\d=0$.
In the non-interacting case, however, the absolute value of $\tilde{I}_3$ is quite small
for the entire parameter regime compared with that in the interacting case. 
This result obviously corresponds the phase diagram of the TRIM, in which only the
localized phase exists.
The state at the IRC point [$\d=0$] is recognized as a particular localized 
state \cite{Kovacs}, and therefore, $\tilde{I}_3$ has a peak at that value. 
The calculated system-size dependence shows that the curves of $\tilde{I}_3$ does not 
cross with each other indicating non-existence of phase transitions for the $g=0$ system
besides $\d=0$.
[However, we shall give a comment on this point at the end of this subsection.]

On the other hand for the interacting case of $g=0.2$, $\tilde{I}_3$ in Fig.~\ref{Fig7}(b)
exhibits clear scaling behavior with respect to the system size.
The curves of $\tilde{I}_3$ cross with each other at two values of $\d$, indicating
the existence of 
two phase transitions such as the PM-MBL $\to$ ETH $\to$ SG-MBL phases as $\d$ increases.
This results is obviously in good agreement with the observation of the half-chain EE in Sec.~II.
Then, we conclude that the TMI is a good indicator of phase transitions.
We have examined $\tilde{I}_3$ for systems of $L=8, \ 10$ in addition to $L=12$  
for $g=0.2$ [not shown] and found
that the system-size dependence of the saturation time is rather weak.
Then, we expect that the TMI can be of practical use for large but finite systems.

Interestingly enough, $\tilde{I}_3$ is \textit{not} symmetric under the transformation
$\d \rightarrow -\d$ outside of the ETH regime indicating breaking of duality in 
the localized phases. 
As discussed in Sec.~III A, we think that this discrepancy of duality
stems from the SPT order and spatial structure of the stabilizer-qubit.
Also, we comment that for the SG--MBL limit (for large $\d$),
the background values of TMI seem to exhibit very clear system size dependence. 
In fact, we observed the values of $I_3/L$ (\textit{not} $\tilde{I}_3/L$)
for SG--MBL limit are almost independent of the system size (not shown). 
This behavior holds also for the CS-MBL as we see later on, being different
from the PM-BML limit with the LIOMs located at a site.
Therefore, we expect that this result indicates the existence 
of bulk size qubits in scrambling process.
We will perform numerical study to verify this expectation in Sec.~III D.

Here, it is appropriate to comment on the above calculations of the TMI and the static
quantities observing localization properties for the $g=0$ case mentioned in 
Sec.~II.A ~\cite{Laflorencie2022}.
In Ref.~\onlinecite{Laflorencie2022},
detailed study on the static half-chain EE and gap ratio for small $g$ indicates
that an ergodic  phase exists for $-\ln 2 < \d <\ln 2$ for $g\to 0$ in the limit $L \to \infty$.
This comes from the avalanche instability of localization~\cite{Roeck2017} 
in the thermodynamic limit.
On the other hand, the TMI in Fig.~\ref{Fig7}(a) does not exhibit ergodic properties
in that parameter region.
The obtained results for the interacting case with $g\neq 0$
obviously show that the TMI is a good indicator for localization.
However, an apparent discrepancy between the static and dynamic quantities, the TMI, exists
for the $g=0$ case.
Unfortunately, we currently do not have a clear understanding of the origin of this
discrepancy.
One possible origin of this discrepancy is a finite-size effect of the observed TMI,
and if so, numerical study of large scale systems beyond exact diagonalization may be
required.
We shall give more comments on it at the end of Sec.~IV.
\subsection{Quench dynamics of bipartite EE: cluster spin chain}

Let us move on the numerical study of the model $\HCS$ in Eq.~(\ref{HCS}).
We found that there are three phases in the system i.e., as the value of $\d$ increases,
CS-MBL $\to$ ETH $\to$ SG-MBL.
Both the SG-MBL and CS-MBL are the localized topological phase with distinct topological feature, 
where in the SG-MBL limit, paired spectrum appears while quartet
spectrum appears in the CS-MBL limit \cite{Bahri,Decker2020}.
Therefore, it is interesting to see how the TMI behaves in these phases
as both SG-MBL and CS-MBL phases have long-range correlations dictated, e.g., 
by loop orders. 
Also the spatial structures of the stabilizer 
in the two MBL regimes for $\d \to \pm \infty$ are different.
Hence, it is expected that the bulk properties of the information spreading are 
different in the two phases.

\begin{figure}[t]
\begin{center} 
\includegraphics[width=7.0cm]{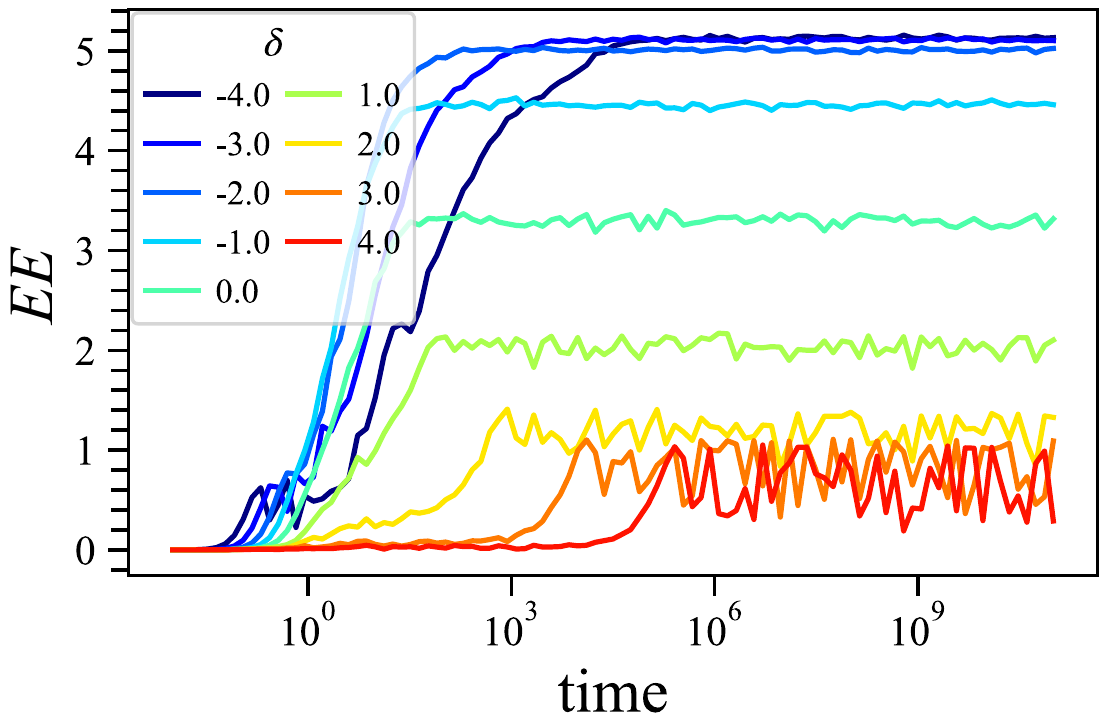}  
\end{center} 
\caption{Quench dynamics of entanglement entropy, $\Et$, in the extended cluster-spin model,
$\HCS$ [Eq.~(\ref{HCS})] with $g=0.2$ and the initial state
$|\ua\ua\ua\cdots\rangle_X$.
$\Et$ increases quite rapidly for $\d=-4, -3$ (the CS-MBL regime), whereas
it does not for $\d=4, 3$ (the SG-MBL regime). The system size is $L=12$}
\label{Fig8}
\end{figure}
In this subsection, we study the quench dynamics of the half-chain EE by varying $\d$.
In Fig.~\ref{Fig8}, the evolution of the EE, $\Et$, is displayed for the initial state
$|\ua\ua\ua\cdots\rangle_X$.
In particular, we are interested in the difference of $\Et$ for the SG-MBL ($\d \gg 1$) 
and CS-MBL ($\d\ll -1$) regimes.

From Fig.~\ref{Fig8}, it is obvious that the system for  $\d=-3, -4$ exhibits large
increases in $\Et$, whereas only small increase for $\d=4, 3$ in the time evolution.
This behavior obviously comes from the difference of the structures of stabilizer-qubit in these two phases,
i.e., the action of the unitary dynamics of $\{\s^x_i\s^x_{i+1}\}$ in the SG-MBL regime
obviously does not induce a significant change of the initial state, i.e., 
quantum information of the initial state does not scramble significantly. 
On the other hand,  $\{\s^x_i\s^z_{i+1}\s^x_{i+2}\}$ in the CS-MBL regime do, 
as the initial state is strongly scrambled by the above the stabilizer-qubits. 
The behavior of $\Et$ depends on the interplay of the stabilizer-qubit and initial state.

Here, we would like to comment: which stabilizer, $\{\s^x_i\s^x_{i+1}\}$ or
$\{\s^x_i\s^z_{i+1}\s^x_{i+2}\}$, dominates is automatically determined by 
parameters of the system under study and system's location properties in the phase diagram.
In other words, study of the phase diagram for the qubit system is required
in order to have stable stabilizers as desired.

In the following subsection, we shall study the TMI, which reflects nature of the time-evolution unitary itself.

\begin{figure}[t]
\begin{center} 
\includegraphics[width=7.0cm]{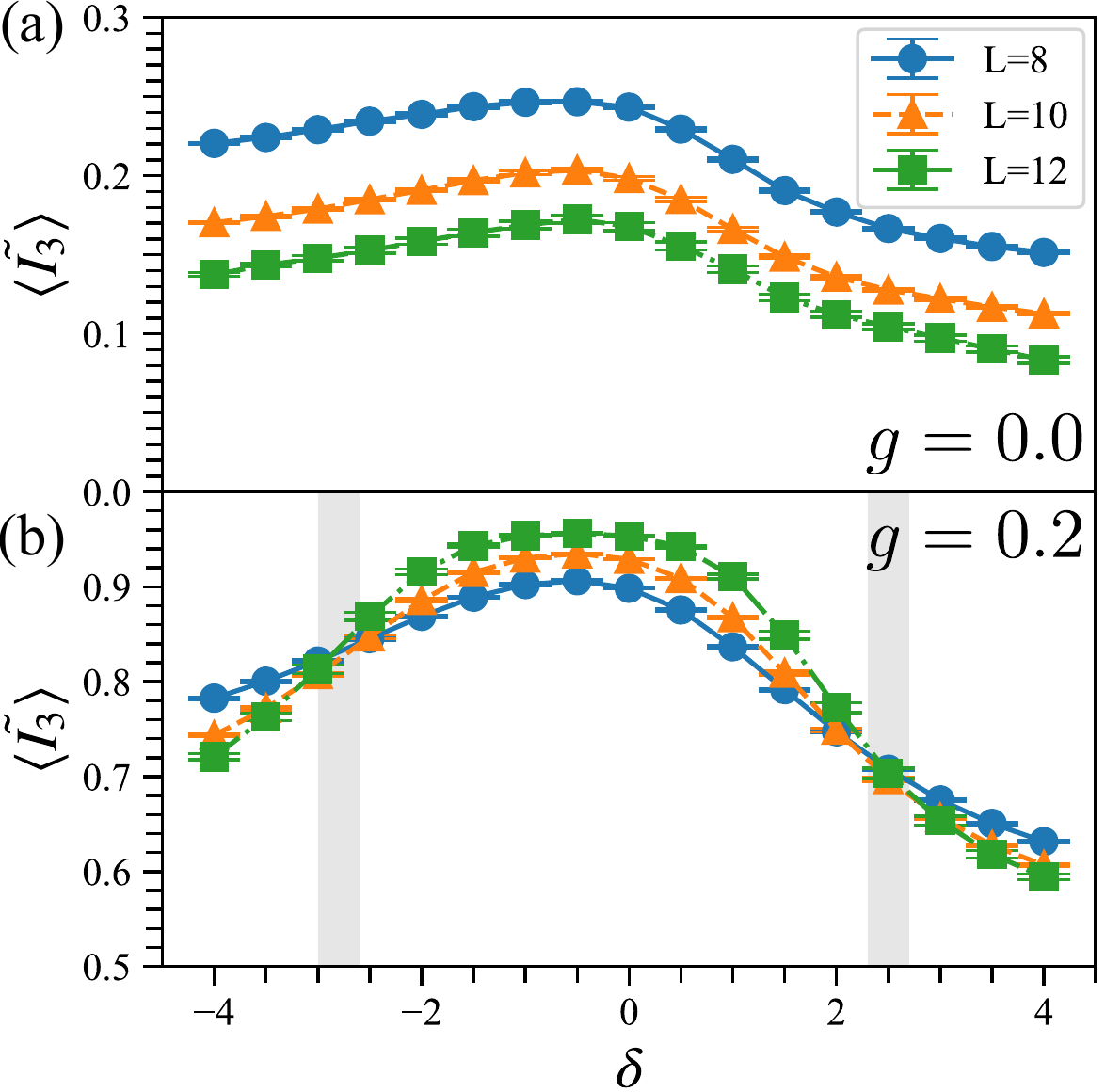}  
\end{center} 
\caption{Saturation values of the TMI for the extended cluster spin model 
in Eq.~(\ref{HCS}) with various system sizes (a) $g=0$ case: 
The data show that $\tilde{I}_3$ is a smooth function of $\d$, there are no crossings
of the curves, and no increase in $\tilde{I}_3$ as $L$ gets larger.
These behaviors indicate a direct transition between the SG-MBL and CS-MBL.
(b) $g=0.2$ case: the curves cross with each other at two phase transition points
observed by the EE. These results were obtained by averaging over the $1000, 500,$ and $100$ disorder realization for L=8, 10, and 12 systems.
}
\label{Fig9}
\end{figure}
\subsection{Tripartite mutual information: cluster spin chain}

In this subsection, we show the calculations of the TMI, $\tilde{I}_3$, for the cluster
spin chain, $\HCS$ in Eq.~(\ref{HCS}). 
In the numerical calculation, we focus on the disorder average of the normalized TMI $\langle \tilde{I}_3\rangle$.
We observed that $\tilde{I}_3$ has a stable time evolution (not shown), and 
in Figs.~\ref{Fig9}(a) and (b), 
we display the saturation values of $\tilde{I}_3$ as a function of $\d$ for various 
system sizes with the $A$ and $B$ ($C$ and $D$) $L/2$-chains.
As in the random Ising spin chain, the data of  $\tilde{I}_3$ for various system 
sizes indicate the existence of two kinds of phase transitions for the $g=0.2$ case,
as indicated by the calculation of the EE in Sec.~II.
As in the Ising spin chain, $\tilde{I}_3 \sim 0.6$ in the SG-MBL regime ($\d\gg 1$).
On the other hand in the CS-MBL ($\d\ll -1$), $\tilde{I}_3$ has a larger value compared
with that value, i.e., $\tilde{I}_3 \sim 0.7$.
From the observation obtained in the investigation of the random Ising spin chain, 
this behavior comes from the difference in the spatial structure of the stabilizers
and the resultant SPT orders. 
On the other hand for the $g=0$ case, the TMI exhibits smooth curves indicating
a direct phase transition between the SG-MBL and CS-MBL 
at least for the small but finite systems.

From the investigation of the TMI given so far, we want to see if there exist some 
other quantities concerning the TMI, which reflects spatial magnitude of logical 
(stabilizer) qubits in the MBL states.
To this end, we calculate the TMI as varying the size of the subsystem $A$
and $D$, that is, changing the partitioning of the in and out Hilbert spaces. 
In particular, we are interested in partitioning with \textit{two-site} $A$ and $D$
(we denote as $L_A=L_D=2$) subsystem in Fig.~\ref{Fig3}.

\begin{figure}[t]
\begin{center} 
\includegraphics[width=7.0cm]{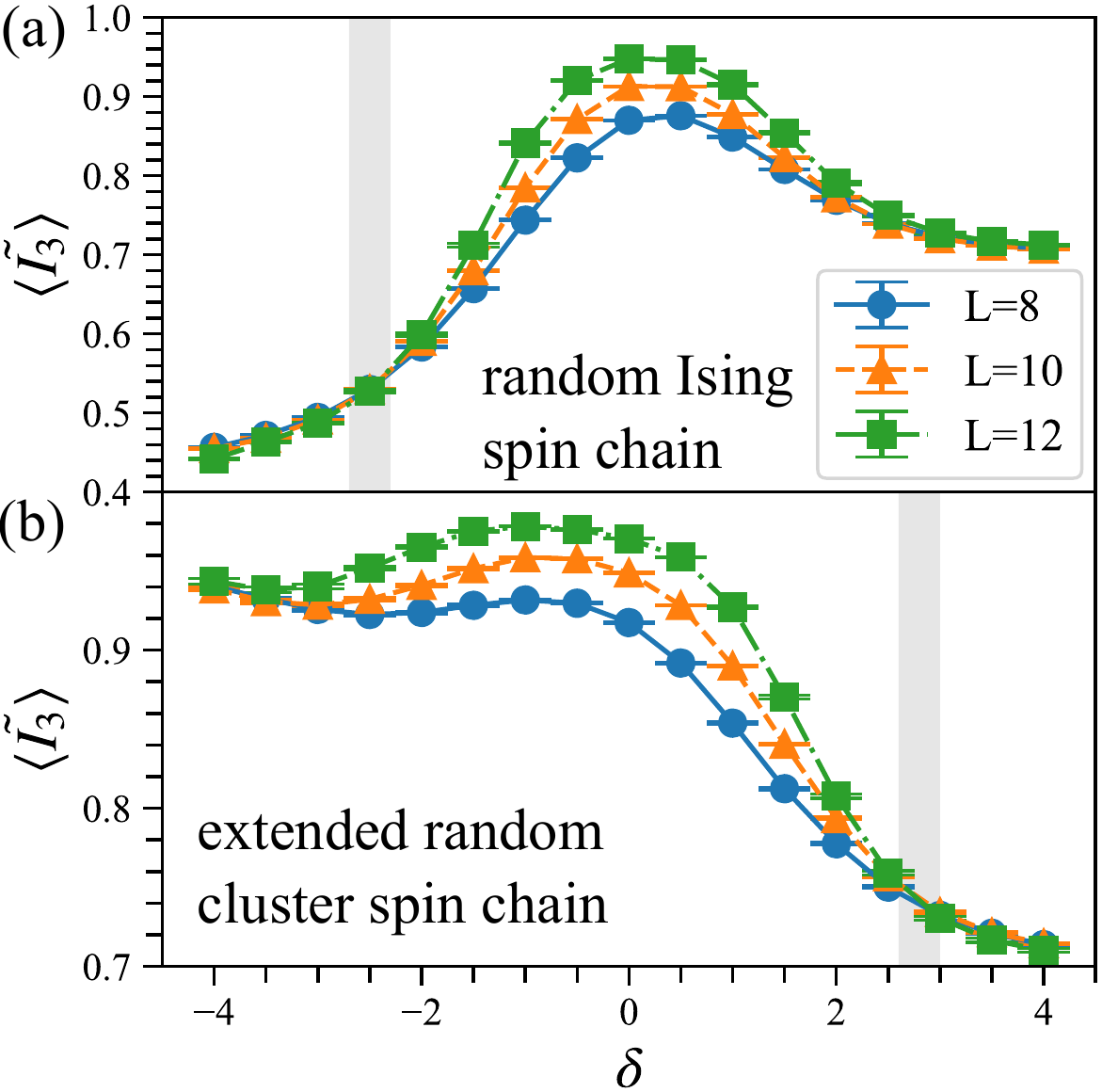}  
\end{center} 
\caption{Saturation values of the TMI for various system sizes under two-site 
partitioning:(a) random Ising spin chain for $g=0.2$.
(b) extended random cluster spin chain for $g=0.2$. These results were obtained by averaging over the $1000, 500,$ and $100$ disorder realization for L=8, 10, and 12 systems.
}
\label{Fig10}
\end{figure}
\begin{figure}[t]
\begin{center} 
\includegraphics[width=7.0cm]{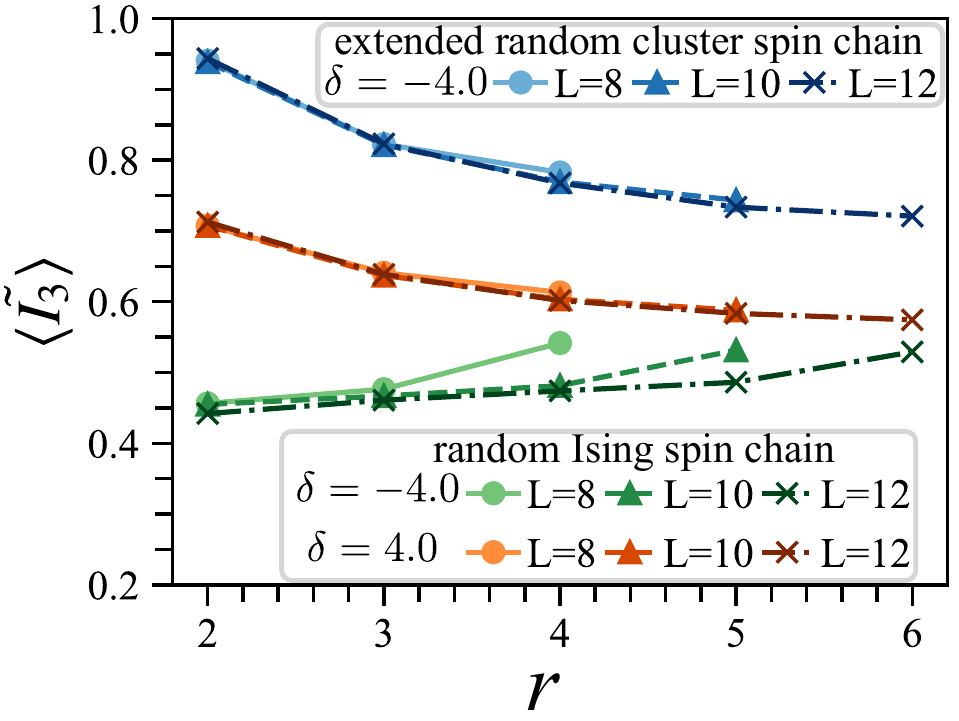}  
\end{center} 
\caption{Saturation values of the TMI for various partitioning of the system.
In the PM-MBL, $\tilde{I}_3$ is an increasing function of $r=L_A=L_D$,
indicating that quantum information encoded initial $A$-system remains
in the $C$-subsystem, as it is expected.
On the other hand for the SG-MBL and CS-MBL with topological order,
$\tilde{I}_3$ is a decreasing function of $r$.
This unexpected result comes from a finite spatial magnitude of stabilizer qubit
in the topological state.
}
\label{Fig11}
\end{figure}

In Figs.~\ref{Fig10}(a) and (b), we show the numerical calculations $\tilde{I}_3$ for 
the random Ising spin chain and extended cluster spin models under the above mentioned
two-site partitioning.
In the SG-MBL phase of the random Ising spin chain and also CS-MBL phase in the 
cluster spin model, 
$\tilde{I}_3$  has larger values compared to those in the equal-length partition.
In particular in the CS-MBL limit, $\tilde{I}_3$ is an increasing function of $|\d|$,
and $\tilde{I}_3\simeq 1.0$, indicating \textit{apparent chaotic behavior} of the CS-MBL for two-site partitioning. 
On the other hand in the PM-MBL state in Fig.~\ref{Fig10}(a), 
$\tilde{I}_3$ is a decreasing function of $|\d|$.
In order to investigate this peculiar behavior of $\tilde{I}_3$, 
we calculate $\tilde{I}_3$ for other partitioning such as $L_A=L_D\equiv r=3, \cdots, L/2-1$ for the deep PM-MBL, SG-MBL and CS-MBL phases.
The results are shown in Fig.~\ref{Fig11}.
We readily find that $\tilde{I}_3$ in the PM-MBL is an increasing function of $r$,
indicating that quantum information encoded in the subsystem $A$ is remaining 
inside of the subsystem $C$ until saturation is achieved in the dynamics.
Contrary to the above plausible result of the PM-MBL, in the SG-MBL
and also CS-MBL with SPT order, $\tilde{I}_3$ is a decreasing function of $r$,
and this decreasing tendency is stronger in the CS-MBL than in the SG-MBL.

It is obvious that the above peculiar phenomenon is related to the spatial structure of the
stabilizer-qubits, i.e., in the deep MBL's regime, $\{\s^x_i\s^x_{i+1}\}$ in the SG-MBL and
$\{\s^x_i\s^z_{i+1}\s^x_{i+2}\}$ in the CS-MBL, 
is equal or larger than two site and also these forms are deformed by the interactions, 
whereas the stabilizer-qubit in the deep 
PM-MBL is a nearly single spin $\{\s^z_i\}$.
[For the stabilizer-qubit in the CS-MBL, please see the following analytical discussion.]
This fact means that quantum information encoded in smaller $A$-subsystem 
than stabilizer-qubits \textit{underflows} a stabilizer-qubit, as information,
which is to be encoded in stabilizer-qubit, is 
lost by tracing out quantum information in the $B$-subsystem.
As a result, $\tilde{I}_3$ exhibits a chaotic-like behavior even in the MBL state.
In other words, the above calculation can exhibit spatial magnitude of LIOMs 
in the MBL regimes.
Obviously, the existence of the stable stabilizer-qubits with a finite magnitude also
supports MBL and the SPT order.
Therefore, the present phenomenon is expected to be rather universal. 

As one may wonder how the stabilizer-qubits (therefore, LIOMs) are deformed (or dressed) by the existence
of other terms in the Hamiltonian besides the mutually commuting terms, let us analyse
the random CS model in Eq.~(\ref{HCS}).
To this end, it is convenient to introduce the following operators,
\be
&& K_i=\s^x_{i-1}\s^z_i\s^x_{i+1},  \nonumber  \\
&& K^{\pm}_i = {1\over 2} (\s^x_i\pm i\s^x_{i-1}\s^y_i\s^x_{i+1}),
\label{Ks}
\ee
and
\be
&(K^+_i)^\dagger=K^-_i, \; (K^\pm_i)^2=0, \;
K^+_iK^-_i+K^-_iK^+_i=1, \nonumber \\
& K^+_iK^-_i = {1 \over 2} + {1\over 2} K_i, \; [K_i,K^\pm_i]=\pm 2 K^\pm_i.
\label{Ks2}
\ee
Therefore, $K^\pm_i$'s are nothing but hard-core bosons, and $K_i$'s are their number operators.
The leading terms of $H_{\rm CS}$ in Eq.(\ref{HCS}), $\{\lambda_i K_i\}$, describe a random potential,
and $\{\s^x_i\s^x_{j}=(K^+_i+K^-_i)(K^+_j+K^-_j)\}$ are hopping terms.
It is not so difficult to show that the other terms in $H_g$, $\{\s^z_i\s^z_{i+1}\}$,
describe local interactions between the hard-core bosons.
A Majorana representation can be introduced straightforwardly by
$$
\chi^1_i\equiv (K^+_i+K^-_i), \ \chi^2_i\equiv {1 \over i}(K^+_i-K^-_i).
$$
By the above observation, the LIOMs are given by $\{K_i\}$'s in the CS-MBL limit, and 
in the deep MBL regime, the hopping  makes $\{K_i\}$'s fluctuate around
their original location, and dressed LIOMs
are local linear combinations of $\{K_i\}$'s as in Anderson localization. 
There, weak interactions by the $g$-terms can be treated perturbatively and 
induce MBL.
Investigation on similar situation to the above for spin systems 
in strong random fields indicates
that the LIOMs are well described by dressed spins with very narrow tail,
very close to physical qubits (spins) \cite{Chandran2015,OPDM2017,OKI2022_CL}.
In the present system, the SPT order exhibits the stability of $\{K_i\}$'s,
as the string order is nothing but the expectation value of a product 
of $\{K_i\}$'s.
Furthermore from the data in Fig.~\ref{Fig11}, we expect  that some fraction of 
stabilizer-qubits have large scale cat-state like nature, which come from
the $\mathbb{Z}_2$-symmetry and SPT order and reflect $\tilde{I}_3$ for $r \sim L/2$ 
in Fig.~\ref{Fig11}.

\section{Discussion and conclusion}

In this paper, we studied two kinds of quantum spin chains, both of which have a nontrivial phase diagram.
By investigating the EE and the order parameters, we first clarified phase diagrams of 
the random Ising spin chain and extended random cluster spin chain.
Then, we studied the quench dynamics of the EE for various initial states to
obtain an intuitive picture of the quantum information spreading in these systems.
We noted the breakdown of duality of the Ising spin chain in the quench dynamics,
which gives us a clue to understand how quantum information scrambles.
Finally, we calculated the TMI and obtained important perspectives on the information spreading.

Findings, which we obtained in this work, are summarized as follows;
\begin{enumerate}
    \item In order to observe the quench dynamics from the view of duality, 
   initial states have to be prepared carefully as they are connected by duality.
   \item In the time evolution, the EE and TMI exhibit oscillating behavior in the TRIM
   because of its Griffiths phase character, whereas they become stable by adding
   interactions $H_g$ in Eq.~(\ref{HIC}).
   \item In quench dynamics of the random Ising spin chain, the EE exhibits small 
   but finite breaking of duality.
   \item The above breaking of duality also emerges in the TMI in the MBL regimes.
   \item The TMI exhibits clear system-size dependence and it is a good indicator 
   for phase transitions, especially for interacting systems.
   \item The time evolution of the TMI is stable in almost all cases except the Griffiths regime, and its saturation values exhibit rather characteristic behavior, in
   particular, in the partitioning of subsystems with unequal length.
\end{enumerate}
 
From the above findings obtained by the numerical calculation, we have got an 
important insight into 
how quantum information in the bulk is encoded in quantum spin chains and how randomness
(disorder) influences quantum information spreading.
Calculation of the TMI in the two-site partitioning of chain reveals that quantum
information is encoded in stabilizer-qubits in the MBL regimes. We also note
that as shown in the numerical result in Fig.10, the spatial structure of the
stabilizer-qubit in the MBL regime is robust for (at least) weak interactions. 
The stabilizer-qubits are nothing but the local-bits or LIOMs, which were introduced
to explain logarithmic time evolution of the quench EE in the MBL regimes.
In the ordinary spin chains in random magnetic fields, the local-bits are described by
dressed Pauli spin operators that substantially reside on a single site in the
localization limit.
On the other hand in the present work, the local-bits are explicitly
given by the multi-site spin composites (stabilizers) such as
$\{\s^x_i\s^x_{i+1}\}$ and
$\{\s^x_i\s^z_{i+1}\s^x_{i+2}\}$ in the MBL limit, and it is expected that they are
spatially expanded by the additional interactions between spins.
Therefore, they behave differently from the ordinary local-bits because of their spatial magnitude, as explicitly observed by the TMI in the two-site partitioning.
In quantum information science, the viewpoint of the spatial structure of
stabilizer-qubits may be important and useful on constructing practical quantum circuit 
by using quantum physical devices.

Another interesting observation obtained in this work concerns the IRC and Griffiths
nature of the random Ising spin chain.
In the TRIM, almost all quantities, observed in this work, exhibit unstable behavior
in the time evolution.
The study of the TRIM has a long history but has not been completed yet.
The present work reveals its peculiar behavior in quantum information aspect.
We think that this finding and detailed study of the TRIM from quantum information viewpoint will uncover the nature of the IRC and Griffiths phase. This is a future problem.

The above observations clearly indicate that nature of quantum information scrambling is
determined by the phase diagram of the model describing 
that quantum system, and stability of stabilizer can be predicted by the knowledge of 
the phase diagram.
This result may be of great importance for, e.g., constructing logical qubit by means of
stabilizer code.
How to utilize the knowledge of phase diagram for construction of logical code, etc. is 
an interesting future problem.
One example in this direction is a random circuit of projective transverse field Ising
model studied in Refs.~\onlinecite{Lang2020,Shengqi,MPAFisher2021}.
In that system, projective measurements (stabilizers), which are given by $\{\s^z_i\}$
and $\{\s^x_i\s^x_{i+1}\}$  (or $\{\s^x_i\}$ and $\{\s^z_i\s^z_{i+1}\}$), are applied in each time step with probability $p$ and $1-p$,
respectively.
It is expected that $p$ plays a role of $\d$ in the random Ising spin chain in this
work, and the random distribution of the stabilizers corresponds to random variables
$\{J_i\}$ and $\{h_i\}$.
In fact, it was observed that the EE tends to $\log 2 \ (\log 1=0)$ for the limit
$p \to 0 \ (1)$ as in the random Ising spin chain. 
Furthermore, a phase transition takes place at $p=0.5$, and the mutual information
has a finite value for $p<0.5$, whereas it vanishes for $p>0.5$.
The phase for $p<0.5$ is regarded as a spin glass phase with finite bond percolation.
These results obviously coincide with the behavior of the SG-MBL and PM-MBL
in the random Ising spin chain.
Similar projective random circuit system corresponding to the XZZX spin chain was 
also studied very recently\cite{Klocke2022}.
Then, it is interesting to study a random circuit of projective measurements
corresponding to the extended random cluster spin chain investigated in this
work. This work is in progress. 
Another direction is to study the relationship between 
TMI and more practical information spreading\cite{Ashhab2015}, and such an application may be interesting.

We also note that there is close connection between the present work and topological
Majorana quantum memory~\cite{Kitaev2001,Bravyi2010,Nahum2020}.
Knowledge of stable quantum storing in that system by topologically produced global
Bell clusters helps us to get an intuitive picture of the significantly large TMI in 
the SG-MBL and CS-MBL regimes.
Detailed study on the relation is a future work.

Finally, we would like to comment on recent studies on MBL transition in the thermodynamic
limit~\cite{Sels2021,Morningstar2022,Sels2021v2}.
These works indicate that `putative' MBL, which is observed in finite systems,
cannot survive in the thermodynamic limit.
Idea named `finite-size MBL regime' was proposed, which is to be distinguished from
the genuine MBL phase.
Most of studies focused on the XXZ and XXX spin models in a random external field,
and therefore, the investigation of the TMI for the XXX spin model in
Ref.~\onlinecite{Bolter2022} is quite useful.
There, the behavior of the TMI was studied by the exact diagonalization in small systems, 
and it exhibits a phase transition-like behavior with a critical magnitude of 
the random field,
which is close to the ones obtained by the gap ratio and the half-chain EE.
This indicates that `MBL phase transition' observed by the TMI may correspond to
the finite-size MBL regime.
The phase transitions observed in this work may be a crossover to the finite-size MBL
regime, in particular, the SPT nature protected by MBL may disappear in the
thermodynamic limit.
However from quantum information point of view, our findings in the present study
are useful as devices in quantum-information instruments are of finite size and 
a period using them is also finite.
Our work clarified the parameter regimes, in which relevant states emerge and are stable,
and gives guides for constructing quantum network using many-body spins such as 
described by cluster spin models. 

\section*{Acknowledgements}
T.O. has been supported by the Program for Developing and Supporting the Next-Generation of Innovative Researchers at Hiroshima University. 
This work is also supported by JSPS KAKEN-HI Grant Number JP21K13849 (Y.K.).


\appendix
\section*{Appendix: Numerical validity of the saturation value of the TMI}
\begin{figure}[t]
\begin{center} 
\includegraphics[width=7.0cm]{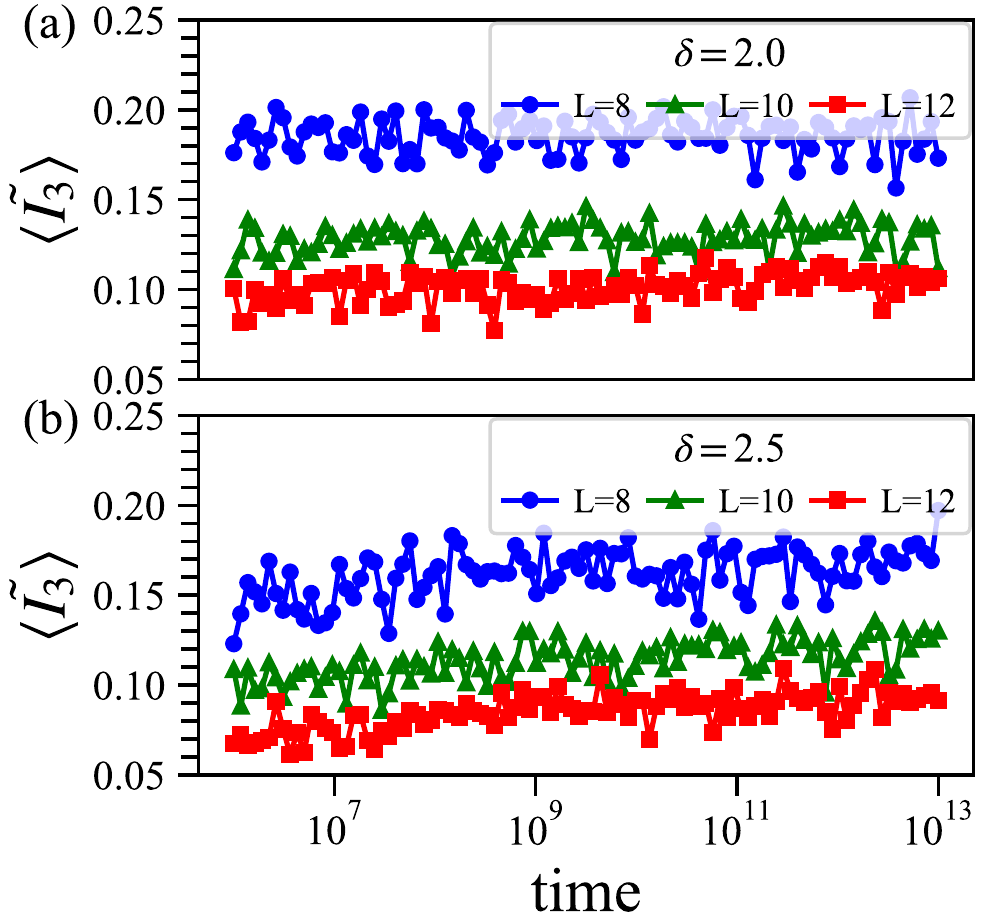}  
\end{center} 
\caption{TMI dynamics of random Ising chain ($g=0$) with various system sizes in the long time limit: (a) $\delta=2.0$ and (b) $\delta=2.5$. We employ the same random coefficients $J_i$ and $h_i$ of Fig. \ref{Fig6}. 
and take the same disorder averages. 
}
\label{Fig12}
\end{figure}
In this paper,
we have found that the saturation value of the $\tilde{I}_3$ distinguishes the phases.
However, in Fig. \ref{Fig6},
one can see $\tilde{I}_3$ with $\delta=2.0$ and $2.5$ do not 
seem  to reach the saturation values.
In this Appendix, 
we verify that such a weakly increasing nature of $\tilde{I}_3$ does not affect the main results.
Figure~\ref{Fig12} shows $\tilde{I}_3$ dynamics in the further long time period up to $t=10^{13}$
with the same numerical conditions as Fig.~\ref{Fig6}. 
$\tilde{I}_3$ may increase with time evolution;
however, as the system size $L$ increases,
$\tilde{I}_3$ decreases, which implies the absence of the crossing for $\tilde{I}_3$,
i.e., weakly increase in $\tilde{I}_3$ does not affect the main results.



\begin{thebibliography}{99}
\bibitem{Haar_ND}
$I^{H}_3$ is calculated under the dimension $2^L$ including full-sectors of total magnetization.
\bibitem{Haar_val}
When all the subsystems are the same length, the values of $I^{H}_3/L$ are $-0.820539$,$-0.855923$, and $-0.879814$ for $L=8$, $10$, and $12$.When the length of subsystems A and D are $2$ and subsystems B and C are $L-2$, the values of $I^{H}_3$ are $-3.3060497$,$-3.309270$, and $-3.308958$ for $L=8$, $10$, and $12$.
\bibitem{EE_long_range}
We note that this statement does not necessarily mean the SG-MBL and CS-MBL phases are long-range entangled phase. 
Both phases are symmetry-protected topological phase with localization nature, that is,
each static eigenstate is short-range entangled, which obeys area-law.
\end{thebibliography}
\end{document}